\documentclass[prd,onecolumn,notitlepage,eqsecnum,nofootinbib,floatfix,superscriptaddress]{revtex4-1}
\usepackage{amsmath, amsthm, amssymb, amsfonts, amsbsy,mathrsfs}

\usepackage{graphicx, epstopdf}
\DeclareGraphicsExtensions{.eps}
\graphicspath{{./images/}}

\usepackage{calligra,bm}
\usepackage{stmaryrd}

\usepackage[hidelinks,bookmarks=true]{hyperref} 
\hypersetup{pdfstartview=FitH,pdfhighlight=/O,colorlinks=false}

\begin{document}

\preprint{UTTG-08-17}

\title{From path integrals to the Wheeler-DeWitt equation: Time evolution in spacetimes with a spatial boundary}

\author{Justin C. Feng}
\author{Richard A. Matzner}
\affiliation{Department of Physics, The University of Texas at Austin, Texas 78712, USA}

\date{\today}

\begin{abstract}
We reexamine the relationship between the path integral and canonical formulation of quantum general relativity. In particular, we present a formal derivation of the Wheeler-DeWitt equation from the path integral for quantum general relativity by way of boundary variations. One feature of this approach is that it does not require an explicit 3+1 splitting of spacetime in the bulk.  For spacetimes with a spatial boundary, we show that the dependence of the transition amplitudes on spatial boundary conditions is determined by a Wheeler-DeWitt equation for the spatial boundary surface. We find that variations in the induced metric at the spatial boundary can be used to describe time evolution---time evolution in quantum general relativity is therefore governed by boundary conditions on the gravitational field at the spatial boundary. We then briefly describe a formalism for computing the dependence of transition amplitudes on spatial boundary conditions. Finally, we argue that for nonsmooth boundaries, meaningful transition amplitudes must depend on boundary conditions at the joint surfaces.
\end{abstract}


\maketitle
\tableofcontents

\section{Introduction}
In this paper, we present a formal derivation of the Wheeler-DeWitt equation from the path integral for quantum general relativity, and extend our formalism to describe the dependence of path integral transition amplitudes on spatial boundary conditions. The reader may be aware of existing derivations of the Wheeler-DeWitt equation in the literature that use the path integral as a starting point \cite{HartleHawking1983,Halliwell1988,HalliwellHartle1991,Barvinsky1993a} (there is also an old paper \cite{Leutwyler1964} that argues that the path integral for quantum general relativity satisfies the Wheeler-DeWitt equation). We also note that the matter of spatial boundary conditions has been addressed before \cite{HaywardWong1992} in the form of a boundary Schr{\"o}dinger equation. The formalism we present in this paper, however, has a feature that is not present in the existing derivations; in particular, it does not require a 3+1 splitting of spacetime in the bulk.\footnote{Our formalism is conceptually similar to that of the general boundary formulation of quantum field theory\cite{Oeckl2003,Oeckl2008} in that we construct transition amplitudes for compact regions of spacetime with a connected boundary.} The absence of this requirement is particularly useful for studying spacetimes with spatial boundaries and allows us to establish the dependence of transition amplitudes on spatial boundary conditions. In particular, we find that time evolution for such transition amplitudes corresponds to variations of the induced 3-metric at the timelike spatial boundary. While the general idea that spatial boundary conditions establish time evolution in quantum gravity may be found in the existing literature,\footnote{See for instance the following papers on the general boundary formalism: Refs. \cite{Oeckl2003b,Oeckl2008}. This idea is discussed in Ref. \cite{Baezetal1995} and worked out for loop quantum gravity (our approach differs in that we work exclusively in the 3-metric representation, and that our starting point is the path integral formalism, rather than the canonical formalism). Also, Smolin Ref. \cite{Smolin1993} points out that the problem of time is avoided in asymptotically flat spacetimes.} the results we present formalize this idea and elucidate the relationship between time evolution and spatial boundaries. Furthermore, our results demonstrate that the boundary Schr{\"o}dinger equation in Ref. \cite{HaywardWong1992} is in fact kinematical; the dependence of the transition amplitudes on spatial boundary conditions is determined by a boundary Hamiltonian constraint on the path integral transition amplitudes---the ``Wheeler-DeWitt'' equation for timelike spatial boundaries.

The derivation we present in this paper is based on the Weiss variational principle \cite{SudarshanCM,MatznerShepleyCM}. In the Weiss variation, the boundaries of the action integral are not held fixed---we include displacements of the boundaries/end points. Though the Weiss variation is rarely referred to as such in the literature, the Weiss variation itself is a well-known result and should be recognizable to readers familiar with the derivation of Noether currents from the action. This paper is organized as follows. We begin by briefly describing the Weiss variation in mechanics and its application to the gravitational action; a more detailed discussion may be found in Ref. \cite{FengMatzner2017a}. We also review the concept of superspace, which will be used to motivate an operator-path integral correspondence. We then review the derivation of the Schr{\"o}dinger equation from the quantum mechanical path integral and discuss correspondence rules between path integrals and operators. Afterward, we present our derivation of the  Wheeler-DeWitt equation for spacetimes \textit{without} a spatial boundary. Finally, we extend our derivation to include transition amplitudes for spacetimes \textit{with} a spatial boundary and discuss the matter of time evolution for such transition amplitudes.

\section{Weiss Variation}
We summarize the Weiss variational principle in mechanics \cite{SudarshanCM,MatznerShepleyCM,Weiss1936} and briefly discuss its application to the gravitational action. A detailed discussion of the Weiss variation and its application to the gravitational action may be found in Ref. \cite{FengMatzner2017a} (also see references contained therein).

\subsection{Mechanics}
Consider a mechanical system with \(N\) degrees of freedom \(q^i \in \mathbb{Q}\), with \(\mathbb{Q}\) being a manifold called the configuration space manifold. The system is described by the action:
\begin{equation} \label{3M-Action}
\begin{aligned}
S[q]:=\int^{t_2}_{t_1} L(q,\dot{q},t) \> dt,
\end{aligned}
\end{equation}

\noindent where the degree of freedom index \(i\) is suppressed in the argument of the Lagrangian, which is assumed to be nondegenerate in \(\dot{q}^i\). In the Weiss variation, we include temporal displacements of the end points of the form:
\begin{equation} \label{3M-WeissTimeParam}
\begin{aligned}
{t^\prime}_1 &=t_1+ \lambda \> \tau_1\\
{t^\prime}_2 &=t_2+ \lambda \> \tau_2,
\end{aligned}
\end{equation}

\noindent where \(\lambda\) is an infinitesimal parameter. We perform variations for \(q^i(t)\),
\begin{equation} \label{3M-WeissPaths}
\begin{aligned}
{q^\prime}^i(t)=q^i(t)+ \lambda \> \eta^i(t),
\end{aligned}
\end{equation}

\noindent where \(\eta^i(t)\) are functions of \(t\) that define the variations. The first-order (in  \(\lambda\)) variation of (\ref{3M-Action}) takes the form:
\begin{equation} \label{3M-VariationAction}
\begin{aligned}
\delta S&:=S[q^\prime]-S[q]=\varepsilon \int^{t_2}_{t_1} \left(\frac{\partial L}{\partial q^i}- \frac{d}{dt} \left(\frac{\partial L}{\partial \dot{q}^i} \right)\right)\eta^i(t) \> dt + \left( \frac{\partial L}{\partial \dot{q}^i} \> \lambda \> \eta^i(t)+  L \> \Delta t\right) \biggr |^{t_2}_{t_1},
\end{aligned}
\end{equation}

\noindent where \(\Delta t|_{t_1}:=\lambda \, \tau_1\) and \(\Delta t|_{t_2}:=\lambda \, \tau_2\). It is convenient to define the \textit{total} change in the end point values for \(q^i\),
\begin{equation} \label{3M-EndpointDisplacementA}
\begin{aligned}
\Delta q^i_1&:={q^\prime}^i({t^\prime}_1)-q^i(t_1)=\lambda (\eta^i(t_1) + \tau_1 \,  \dot{q}^i(t_1))+O(\lambda^2)\\
\Delta q^i_2&:={q^\prime}^i({t^\prime}_2)-q^i(t_2)=\lambda (\eta^i(t_2) + \tau_2 \, \dot{q}^i(t_2))+O(\lambda^2),
\end{aligned}
\end{equation}

\noindent and also the quantities
\begin{equation} \label{3M-WeissMomDefn}
\begin{aligned}
p_i&:=\frac{\partial L}{\partial \dot{q}^i}\\
\end{aligned}
\end{equation}
\begin{equation} \label{3M-WeissHamDefn}
\begin{aligned}
H:=p_i \> \dot{q}^i-L,
\end{aligned}
\end{equation}

\noindent which the reader may recognize as the conjugate momentum and Hamiltonian. From the above definitions, it is not difficult to rewrite (\ref{3M-VariationAction}) in the following form: 
\begin{equation} \label{3M-WeissVariation}
\begin{aligned}
\delta S &=\lambda \int^{t_2}_{t_1} \left(\frac{\partial L}{\partial q^i} - \frac{d}{dt} \left(\frac{\partial L}{\partial \dot{q}^i} \right)\right)\eta^i(t) \> dt + \left( p_i \> \Delta q^i - H \, \Delta t\right) \biggr |^{t_2}_{t_1}.
\end{aligned}
\end{equation}

\noindent The above expression for \(\delta S\) is the \textit{Weiss variation} of the action. The Weiss variational principle demands that physical paths be described by functions \(q^i(t)\) for which \(\delta S\) consists exclusively of end point/boundary terms, which is equivalent to the statement that the integral over \(t\) in (\ref{3M-WeissVariation}) vanishes; it follows that the Weiss variational principle implies the Euler-Lagrange equations. 

\subsection{Gravitational action without spatial boundary}
We assume a four-dimensional spacetime manifold \(\mathcal{M}\) and use the Misner-Thorne-Wheeler \cite{MTW} signature \((-,+,+,+)\) for the metric tensor \(g_{\mu \nu}\). Greek indices refer to coordinates on the spacetime manifold \(\mathcal{M}\) and its subsets, which we denote as \(\textbf{U}\) or \(\textbf{W}\); \(\textbf{U}\) will be used to indicate a spacetime region without a spatial boundary, and later on, \(\textbf{W}\) will be used to denote a spacetime region with a spatial boundary. Coordinates on \(\mathcal{M}\), \(\textbf{U}\), and \(\textbf{W}\) will be denoted \(x^\mu\) with \(x^0=t\) being the time coordinate. Lowercase latin indices refer to either mechanical degrees of freedom or coordinates on hypersurfaces---the distinction should be apparent from the context. Capital latin indices from the beginning of the alphabet will refer to two-dimensional surfaces in \(\mathcal{M}\).

For simplicity, we first consider a globally hyperbolic spacetime without a spatial boundary; in particular, we require \(\mathcal{M}\) to be a Lorentzian manifold with the topology \(\mathbb{R} \times \Sigma\), where \(t \in \mathbb{R}\) is a time coordinate and \(\Sigma\) is a three-dimensional manifold without a boundary. We then consider a subset \(\textbf{U} \subset \mathcal{M}\) with boundary surfaces \(\Sigma_{I}\) and \(\Sigma_{F}\), which need not be surfaces of constant \(t\). We require the surfaces \(\Sigma_{I}\) and \(\Sigma_{F}\) to be spacelike, with a positive-definite induced metric \(\gamma_{ij}\). The gravitational action over \(\textbf{U}\) may then be written as
\begin{equation} \label{3GR-GravitationalAction}
\begin{aligned}
S_{GR}[g^{\mu \nu}]=\frac{1}{2 \kappa}\int_{\textbf{U}} \> R \>\sqrt{|g|} \> d^4 x+\frac{1}{\kappa}\int_{\partial \textbf{U}} K \> \varepsilon \> \sqrt{|\gamma|}  \> d^3 y ,
\end{aligned}
\end{equation}

\noindent where \(x^\mu\) denote coordinates on the spacetime manifold \(\mathcal{M}\), \(R\) is the Ricci curvature scalar of \(\textbf{U}\), \(K\) is the mean curvature of the boundary surface \(\partial \textbf{U}=\Sigma_I \cup \Sigma_F\), and we have defined \(\gamma:=\det(\gamma_{ij})\) (with \(\gamma_{ij}\)) and \(\varepsilon:=n^\mu \, n_\mu=\pm 1\), where \(n^\mu\) is the unit normal vector to \(\partial \textbf{U}\) (here, \(\varepsilon=-1\)). The quantity \(\kappa=8 \pi G\), where \(G\) is Newton's gravitational constant.

The Weiss variation includes boundary displacements. To characterize boundary displacements, we begin by placing coordinates \(y^i\) on the boundary \(\partial \textbf{U}\). The boundary may be parametrically defined by functions \(x^\mu (y)\), which specify the position of the boundary \(\partial \textbf{U}\) in \(\mathcal{M}\). The induced metric \(\gamma_{ij}\) on the boundary may then be written as:
\begin{equation} \label{3GR-BoundaryInducedMetric}
\gamma_{ij}=\frac{\partial x^\mu}{\partial y^i}\frac{\partial x^\nu}{\partial y^j} \, g_{\mu \nu},
\end{equation}

\noindent and it is well known that the induced metric takes the form \(\gamma_{\mu \nu}=g_{\mu \nu}- \varepsilon \, n_\mu \, n_\nu\) in the bulk coordinate basis. Displacements of the boundary in the manifold \(\mathcal{M}\) may be characterized by adding to  \(x^\mu (y)\) a function \(\delta x^\mu (y)\) so that the displaced boundary may be parametrically defined by the functions:
\begin{equation} \label{3GR-BoundaryDisplacement}
x^{\prime \mu} (y)=x^\mu(y)+\delta x^\mu(y).
\end{equation}

The presence of the boundary term (called the Gibbons-Hawking-York boundary term) in the gravitational action \(S_{GR}[g^{\mu \nu}]\) complicates the derivation of the Weiss variation of \(S_{GR}[g^{\mu \nu}]\). One may obtain the variation of the boundary term by recognizing that the boundary term is the first variation of the area formula; the variation of the boundary term is then given by the second variation of the area formula. We shall not reproduce the derivation here; we refer the reader to \cite{FengMatzner2017a} for the full details of the derivation [a partial justification for (\ref{3GR-GravitationalActionVariationFullWeiss}) is given in the Appendix]. The Weiss variation of the gravitational action takes the form
\begin{equation} \label{3GR-GravitationalActionVariationFullWeiss}
\begin{aligned}
\delta S_{GR}&=\frac{1}{2 \kappa}\int_{\textbf{U}}  \, {G}_{\mu \nu} \,  \delta g^{\mu \nu} \sqrt{|g|} \,  d^4 x+ \frac{ \varepsilon}{2\kappa}\int_{\partial \textbf{U}} \biggl(p_{ij} \, \Delta \gamma^{ij} + \left[ n_\mu \left({^3}{R}-\varepsilon(K^2-K_{ij} \> K^{ij})\right)- 2 \, D_\alpha {p}^{\alpha \beta} \, \gamma_{\mu \beta} \right]\delta x^\mu \biggr) \sqrt{|\gamma|} \, d^3 y,
\end{aligned}
\end{equation}

\noindent where \({G}_{\mu \nu}:={R}_{\mu \nu}-\tfrac{1}{2} \, R \, {g}_{\mu \nu} \) is the Einstein tensor, \(^3R\) and \(K_{ij}\) are the respective Ricci scalar and extrinsic curvature tensor for the boundary \(\partial \textbf{U}\), and \(D_\mu\) denotes the covariant derivative on a hypersurface---in this case, \(D_\mu\) is the covariant derivative on \(\partial \textbf{U}\). We define the following quantity in the basis of the bulk coordinates \(x^\mu\) and the basis of the boundary coordinates \(y^i\),
\begin{equation} \label{3GR-ConjugateMomentumTensor}
\begin{aligned}
p_{\mu \nu}&:=K_{\mu \nu} - K \> \gamma_{\mu \nu}\\
p_{ij}&:=\frac{\partial x^\mu}{\partial y^i}\frac{\partial x^\nu}{\partial y^j}(K_{\mu \nu} - K \> \gamma_{\mu \nu}),
\end{aligned}
\end{equation}

\noindent where \(K_{\mu \nu}\) is the extrinsic curvature tensor in the bulk coordinate basis (for \(K_{\mu \nu}\) and \(K=g^{\mu \nu}\, K_{\mu \nu}\), we use the sign convention found in Refs. \cite{Poisson,Wald}). These two expressions for the tensor \(p_{\mu \nu}\) are equivalent because the tensor \(p_{\mu \nu}\) is tangent to the boundary surfaces, or that \(p_{\mu \nu} n^\mu=p_{\nu \mu} n^\mu=0\). For later use, we define the quantity:
\begin{equation} \label{3GR-ConjugateMomentum}
\begin{aligned}
{P}_{\mu \nu}&:=\frac{\varepsilon}{2 \kappa}\, {p}_{\mu \nu} \, \sqrt{|\gamma|}\\
P_{ij}&:=\frac{\varepsilon}{2 \kappa}\, p_{ij} \, \sqrt{|\gamma|}.
\end{aligned}
\end{equation}

\noindent The quantity \(\Delta \gamma^{ij}\) is the total change in the induced metric at the boundary surface, in the same way that \(\Delta q^i\) is the total change in the variable \(q^i\) at the end points [cf. Eq. (\ref{3M-EndpointDisplacementA})].

We stress that despite the appearance of a hypersurface Ricci scalar \(^3R\) and the extrinsic curvature tensor \(K_{ij}\), the Weiss variation of the gravitational action (\ref{3GR-GravitationalActionVariationFullWeiss}) makes no reference to a 3+1 split in the bulk manifold \(\textbf{U}\); in another paper \cite{FengMatzner2017a}, we show that Eq. (\ref{3GR-GravitationalActionVariationFullWeiss}) may indeed be derived without performing a 3+1 split in the bulk manifold \(\textbf{U}\). As stated before, the hypersurfaces \(\Sigma_{I}\) and \(\Sigma_{F}\) that form the boundary \(\partial \textbf{U}\) need not be surfaces of constant \(t\). As a result, the Ricci scalar \(^3R\) and the extrinsic curvature tensor \(K_{ij}\) are properties of the boundary surfaces \(\Sigma_{I}\) and \(\Sigma_{F}\) and do not necessarily correspond to the surfaces of constant time coordinate \(t\) in the bulk manifold \(\textbf{U}\). 

We note that if the vacuum Einstein field equations \(G_{\mu \nu}=0\) are satisfied, then the vacuum Hamiltonian and momentum constraints take the form
\begin{equation} \label{3GR-HamConstraint}
\begin{aligned}
&2 \, G_{\mu \nu} \, n^\mu \, n^\nu=-\varepsilon({^3}{R} - \varepsilon (K^2 - K_{ij} \> K^{ij}))=0\\
\end{aligned}
\end{equation}
\begin{equation} \label{3GR-MomConstraint}
\begin{aligned}
&\gamma^{\mu \beta} \, G_{\mu \nu}\, n^\nu=D_\alpha p^{\alpha \beta}=0 \>\>\>\>\>\> \Rightarrow \>\>\>\>\>\> \gamma^{ik} \, D_k p_{ij}=0.
\end{aligned}
\end{equation}

\noindent The above constraints in turn suggest that the term proportional to \(\delta x^\mu\) in the variation \(\delta S_{GR}\) (\ref{3GR-GravitationalActionVariationFullWeiss}) vanishes. This result suggests that the Hamiltonian and the canonical energy-momentum ``tensor'' for general relativity vanish on vacuum solutions of the Einstein field equations \cite{FengMatzner2017a}. 

Without loss of generality, we may choose the boundary displacement \(\delta x^\mu(y)\) to be proportional to the unit normal vector \(n^\mu\):
\begin{equation} \label{4QG-NormalBoundaryDisplacement}
\begin{aligned}
\delta x^\mu (y)=n^\mu \, \Delta \tau(y).
\end{aligned}
\end{equation}

\noindent This is because the portion of \(\delta x^\mu(y)\) tangent to the hypersurface corresponds to infinitesimal diffeomorphisms on the boundary surface \(\partial \textbf{U}\). We interpret the quantity \(\Delta \tau (y)\) as the amount (measured in proper time) by which the boundary \(\partial \textbf{U}\) is displaced in the normal direction. The variation (\ref{3GR-GravitationalActionVariationFullWeiss}) simplifies to
\begin{equation} \label{3GR-GravitationalActionVariationFullWeissB}
\begin{aligned}
\delta S_{GR}&=\frac{1}{2 \kappa}\int_{\textbf{U}} \left({R}_{\mu \nu} -  \frac{1}{2} \>R \> g_{\mu \nu} \right) \delta g^{\mu \nu} \sqrt{|g|} \>  d^4 x+ \int_{\partial \textbf{U}} \biggl(P_{ij} \, \Delta \gamma^{ij} -\mathscr{H}_{gf}\,\Delta \tau \biggr) \> d^3 y.
\end{aligned}
\end{equation}

\noindent where we make use of the expression \(\gamma_{\mu \beta}\, n^\mu=0\), and we define the ``gauge-fixed'' Hamiltonian density
\begin{equation} \label{3GR-HamiltonianDensityGaugeFixed}
\begin{aligned}
\mathscr{H}_{gf}(P_{ij},\gamma^{ij}):=-\frac{1}{2\kappa} \, \biggl[{^3}{R}-\varepsilon\, (K^2-K_{ij} \> K^{ij}) \biggr]\sqrt{|\gamma|},
\end{aligned}
\end{equation}

\noindent where \(K_{ij}\) and \(K\) depend on \(P_{ij}\) via the expressions
\begin{equation} \label{3GR-ExtrinsicCurvatureInverted}
\begin{aligned}
K_{ij}=\frac{2\, \kappa \, \varepsilon}{\sqrt{\gamma}} \left({P}_{ij} - \frac{1}{2} \, {\gamma}_{ij} \, {\gamma}^{kl} \, {P}_{kl}\right) & \>\>\>\>\>\>\>\>\>\>\>\>\>\>\>\>\>\>\>\> & K=-\frac{\kappa \, \varepsilon}{\sqrt{\gamma}} \,  {P}_{ij} \, {\gamma}^{ij},
\end{aligned}
\end{equation}

\noindent which can be easily obtained from (\ref{3GR-ConjugateMomentum}). Finally, we integrate (\ref{3GR-HamiltonianDensityGaugeFixed}) to obtain the gravitational Hamiltonian:
\begin{equation} \label{3GR-HamiltonianGaugeFixed}
\begin{aligned}
{H}_{GR}[P_{ij},\gamma^{ij}]&=\int_{\textbf{U}}\mathscr{H}_{gf}(P_{ij},\gamma^{ij}) \, d^3y=-\frac{1}{2\kappa}\int_{\textbf{U}} \, \biggl[{^3}{R}-\varepsilon\, (K^2-K_{ij} \> K^{ij}) \biggr]\sqrt{|\gamma|}\, d^3 y.
\end{aligned}
\end{equation}

\subsection{Superspace: Rewriting the gravitational Hamiltonian}
We now take the opportunity to briefly motivate and review the concept of superspace,\footnote{Not to be confused with the coordinate space of a supermanifold (usually discussed in the context of supersymmetry), which is also called ``superspace.'' Recall that supermanifolds possess both commuting (bosonic) and anticommuting (fermionic/Grassmann-valued) coordinates. To avoid confusion, we propose referring to the coordinate space of a supermanifold as \textit{Grassmann superspace} and \(\mathscr{S}(\Sigma)\) as \textit{Riemannian superspace} when a distinction is needed (we shall not do this outside this footnote).} the space of Riemannian 3-geometries \cite{DeWitt1967,Wheeler1968,Fischer1970Superspace,DeWitt1970GeodesicSheaf} (also see \cite{Giulini2009} and references therein). The concept of superspace will be useful for us because it provides a formalism for general relativity that resembles particle mechanics. We begin by using (\ref{3GR-ExtrinsicCurvatureInverted}) to rewrite (\ref{3GR-HamiltonianDensityGaugeFixed}) as
\begin{equation} \label{3GR-HamiltonianDensityGaugeFixed2}
\begin{aligned}
\mathscr{H}_{gf}(P_{ij},\gamma^{ij})&=-\frac{\varepsilon}{2 \kappa} \, P_{ij} \, G^{ijkl} \, P_{kl} -\frac{1}{2\kappa} \,{^3}{R} \, \sqrt{|\gamma|},
\end{aligned}
\end{equation}

\noindent where we define the following: 
\begin{equation} \label{3GR-DeWittSupermetricTensor}
\begin{aligned}
&G^{ijkl}:=\frac{2 \, \kappa^2}{\sqrt{|\gamma|}} \left (\gamma^{ik} \, \gamma^{jl} + \gamma^{il} \, \gamma^{jk} - \gamma^{ij} \, \gamma^{kl}\right)\\
&G_{ijkl} := \frac{ \sqrt{|\gamma|} }{8 \, \kappa^2}\left( \gamma_{ik}\, \gamma_{jl} + \gamma_{il}\, \gamma_{jk} - 2 \, \gamma_{ij}\, \gamma_{kl}\right).
\end{aligned}
\end{equation}

\noindent The tensor \(G_{ijkl}\) is constructed to satisfy the following property:
\begin{equation} \label{3GR-DeWittSupermetricTensorInv}
\begin{aligned}
&G_{ijab} \, G^{abkl}=\frac{1}{2} \left(\delta^k_i \, \delta^l_j + \delta^l_i \, \delta^k_j\right).
\end{aligned}
\end{equation}

\noindent  We may then construct a ``supermetric'' from the expression \(G_{ijkl}\)\footnote{The supermetric defined above differs from the DeWitt supermetric \cite{DeWitt1967,DeWitt1970GeodesicSheaf} by a factor of \(|\gamma|/4\); this is due to our convention that we used the \textit{inverse} 3-metric \(\gamma^{ij}\), rather than the metric \(\gamma_{ij}\), as coordinates on superspace [we interpret \(\gamma_{ij}=\gamma_{ij}(y,\gamma^{kl}(y))\).].}:
\begin{equation} \label{3GR-SuperspaceSupermetric}
\begin{aligned}
\mathscr{G}_{ij:kl}(y,y^\prime):=\frac{1}{2} \left(G_{ijkl}(y) \, \delta^3(y^\prime-y)+G_{ijkl}(y^\prime) \, \delta^3(y-y^\prime)\right).
\end{aligned}
\end{equation}

\noindent The supermetric \(\mathscr{G}_{ij:kl}(y,y^\prime)\) may be regarded as a metric on the space of Riemannian inverse 3-metric fields \(\gamma^{ij}(y)\) and \(\gamma^{ij}{^\prime}(y)\), which we call \(iRiem(\Sigma)\). In particular, \(\mathscr{G}_{ij:kl}(y,y^\prime)\) defines an inner product for tangent vectors \(\dot{\gamma}^{ij}(y)\) of \(iRiem(\Sigma)\). Now, consider two inverse 3-metrics \(\gamma^{ij}(y)\) and \(\tilde{\gamma}^{ij}(y)\) that are related by coordinate transformations; we observe \(\gamma^{ij}(y)\) and \(\tilde{\gamma}^{ij}(y)\) correspond to distinct points in \(iRiem(\Sigma)\). A more physically relevant concept is that of superspace, which is the space of coordinate-independent Riemannian 3-\textit{geometries} (as opposed to coordinate-dependent inverse 3-metrics). Given a manifold \(\Sigma\), one may define superspace \(\mathscr{S}(\Sigma)\) by the formal construction
\begin{equation} \label{3GR-Superspace}
\begin{aligned}
\mathscr{S}(\Sigma)=\frac{iRiem(\Sigma)}{Diff(\Sigma)},
\end{aligned}
\end{equation}

\noindent in which we mod out \(iRiem(\Sigma)\) by the space of diffeomorphisms [denoted \(Diff(\Sigma)\)] on \(\Sigma\). 

While superspace \(\mathscr{S}(\Sigma)\) is a more physically relevant concept than \(iRiem(\Sigma)\), the difficulty with using \(\mathscr{S}(\Sigma)\) rather than \(iRiem(\Sigma)\) lies in the fact that \(\mathscr{S}(\Sigma)\) is not a manifold. In particular, it has been pointed out that the dimensionality of \(\mathscr{S}(\Sigma)\) can change at points corresponding to geometries that possess a high degree of symmetry \cite{Fischer1970Superspace,DeWitt1970GeodesicSheaf}. It was (nonrigorously) argued in Ref. \cite{DeWitt1970GeodesicSheaf} that one may nevertheless extend \(\mathscr{S}(\Sigma)\) to obtain a manifold which is referred to as \textit{extended superspace} \(\mathscr{S}_{ex}(\Sigma)\). It is beyond the scope of this article to review this matter any further (we refer the reader to Refs. \cite{Fischer1970Superspace}, \cite{DeWitt1970GeodesicSheaf}, \cite{Fischer1986}, and \cite{Giulini2009} for further discussion); for our purposes, it suffices to assume that there exists a manifold \(\mathscr{S}_{ex}(\Sigma)\) that contains \(\mathscr{S}(\Sigma)\) as a subset and admits a surjection (a map that is onto) from \(\mathscr{S}_{ex}(\Sigma)\) to \(\mathscr{S}(\Sigma)\).

Since \(\Sigma\) is assumed to be compact, we may require  \(\mathscr{S}_{ex}(\Sigma)\) to have countable (though infinite) dimension. This requirement is motivated by the observation that functions defined on certain compact manifolds, the $n$-torus \(\mathbb{T}^n\) and the $n$-sphere \(\mathbb{S}^n\) for instance, admit a complete countable basis for functions defined on them (the respective discrete Fourier basis and $n$-spherical harmonic basis) so that the function spaces on \(\mathbb{T}^n\) and \(\mathbb{S}^n\) have countable dimension. If the manifold \(\Sigma\) has a boundary, we impose the appropriate boundary conditions to ensure that the function spaces have countable dimension. If \(\mathscr{S}_{ex}(\Sigma)\) has countable dimension, then given coordinates \(\xi^{\mathfrak{a}}\) on \(\mathscr{S}_{ex}(\Sigma)\), and a map \(\gamma^{ij}(y,\xi):\mathscr{S}_{ex}(\Sigma) \rightarrow iRiem(\Sigma)\), we may formally write the supermetric \(\mathscr{G}^{ij:kl}(y,y^\prime)\) in the coordinate basis on \(\mathscr{S}_{ex}(\Sigma)\) in the following manner:
\begin{equation} \label{3GR-SuperspaceCoordMetric}
\begin{aligned}
\mathscr{G}_{\mathfrak{a} \mathfrak{b}}:=\int_{\Sigma}\left( \int_{\Sigma} \frac{\partial \gamma^{ij}(y,\xi)}{\partial \xi^{\mathfrak{a}}} \, \frac{\partial \gamma^{kl}(y^\prime,\xi)}{\partial \xi^{\mathfrak{b}}} \, \mathscr{G}_{ij:kl} (y,y^\prime) \, d^3y \right) d^3y^\prime.
\end{aligned}
\end{equation}

\noindent Note that, since \(\mathscr{S}_{ex}(\Sigma)\) is infinite dimensional (again, we assume that the dimension is countably infinite), there is no upper bound on the values of the indices \(\mathfrak{a},\mathfrak{b} \in \mathbb{N}\) (\(\mathbb{N}\) being the set of natural numbers).

Given the superspace metric \(\mathscr{G}_{\mathfrak{a} \mathfrak{b}}\), we may regard \(\mathscr{S}_{ex}(\Sigma)\) as an infinite-dimensional pseudo-Riemannian manifold,\footnote{That \(\mathscr{S}_{ex}(\Sigma)\) is a pseudo-Riemannian manifold can be seen by noting that for the special case \(\gamma_{ij}=\delta_{ij}\), the independent components of \(G_{ijkl}\) have five positive roots and one negative root \cite{DeWitt1967}.} and since \(\mathscr{S}_{ex}(\Sigma)\) is of countable dimension, the standard formulas of Riemannian geometry apply to quantities defined on \(\mathscr{S}_{ex}(\Sigma)\). We may, for instance, define an inverse superspace metric \(\mathscr{G}^{\mathfrak{a} \mathfrak{b}}\) that satisfies the condition
\begin{equation} \label{3GR-SuperspaceMetricInverseCondition}
\sum^\infty_{\mathfrak{t}=1} \mathscr{G}^{\mathfrak{a} \mathfrak{t}} \, \mathscr{G}_{\mathfrak{t} \mathfrak{b}}=\delta^{\mathfrak{a}}_{\mathfrak{b}}
\end{equation}

\noindent and the connection coefficients
\begin{equation} \label{3GR-SuperspaceConnection}
\Gamma^{\mathfrak{c}}_{\mathfrak{a} \mathfrak{b}}:= \frac{1}{2} \sum^\infty_{\mathfrak{t}=1}\mathscr{G}^{\mathfrak{c} \mathfrak{t}}\left(\frac{\partial \mathscr{G}_{\mathfrak{t} \mathfrak{b}}}{\partial \xi^{\mathfrak{a}}} + \frac{\partial \mathscr{G}_{\mathfrak{a} \mathfrak{t}}}{\partial \xi^{\mathfrak{b}}} - \frac{\partial \mathscr{G}_{\mathfrak{a} \mathfrak{b}}}{\partial \xi^{\mathfrak{t}}}\right),
\end{equation}

\noindent from which we may construct covariant derivatives on \(\mathscr{S}_{ex}(\Sigma)\). 

These definitions, combined with (\ref{3GR-HamiltonianDensityGaugeFixed2}), permit a rewriting of the gravitational Hamiltonian \(H_{GR}\) in the form:
\begin{equation} \label{3GR-HamiltonianSuperspaceRep}
\begin{aligned}
{H}_{GR}(P_{\mathfrak{a}},\xi^{\mathfrak{a}})&=-\frac{\varepsilon}{2 \kappa} \left( \sum^\infty_{\mathfrak{a}=1}\sum^\infty_{\mathfrak{b}=1} \mathscr{G}^{\mathfrak{a} \mathfrak{b}} \, P_{\mathfrak{a}} \,  P_{\mathfrak{b}}\right) + \Phi(\xi),
\end{aligned}
\end{equation}

\noindent where
\begin{equation} \label{3GR-HamiltonianPotential}
\begin{aligned}
\Phi(\xi):=-\frac{1}{2\kappa} \int_{\Sigma}\,{^3}{R} \, \sqrt{|\gamma|} \, d^3y,
\end{aligned}
\end{equation}

\noindent with \(\gamma_{ij}=\gamma_{ij}(y,\xi)\) and \({^3}{R}={^3}{R}(y,\xi)\). The integral over \(y^i\) in the second term ensures that \(\Phi(\xi)\) is strictly a function of the superspace coordinate \(\xi^{\mathfrak{a}}\).

We note that the Hamiltonian \({H}_{GR}\) (\ref{3GR-HamiltonianSuperspaceRep}) resembles the Hamiltonian of particle mechanics on a Riemannian manifold. A typical Hamiltonian for such a system takes the following form
\begin{equation} \label{3QM-RiemHamiltonian}
H(p,q) = \frac{1}{2 m} \, g^{ij} \, p_i \, p_j + V(q),
\end{equation}

\noindent where \(g_{ij}\) is a Riemannian metric on the configuration space \(\mathbb{Q}\) and \(V(q)\) is the potential. A comparison of the Hamiltonian \({H}_{GR}\) (\ref{3GR-HamiltonianSuperspaceRep}) with the particle Hamiltonian (\ref{3QM-RiemHamiltonian}) suggests that the first term containing the conjugate momenta is a kinetic term
\begin{equation} \label{3GR-KineticTerms}
\frac{1}{2 m} \, g^{ij} \, p_i \, p_j \> \> \leftrightarrow \> \> \frac{1}{2 \kappa} \left( \sum^\infty_{\mathfrak{a}=1}\sum^\infty_{\mathfrak{b}=1} \mathscr{G}^{\mathfrak{a} \mathfrak{b}} \, P_{\mathfrak{a}} \,  P_{\mathfrak{b}}\right),
\end{equation}

\noindent and that the function \(\Phi(\xi)\) is a potential. Of course, the idea that general relativity can be recast as a problem of particle motion in an infinite-dimensional manifold is not new \cite{DeWitt1970GeodesicSheaf,Gowdy1970}.\footnote{In fact, Ref. \cite{DeWitt1970GeodesicSheaf} goes further; the Hamiltonian constraint was used to show that solutions of Einstein's field equations can (with appropriate gauge conditions) be interpreted as geodesics in superspace.} However, the formalism developed in this section will be useful for motivating a correspondence between path integral and operator expressions in quantum general relativity, which we will present in later sections.

\subsection{Gravitational action with spatial boundary} \label{Section:Gravitational Action_With_Spatial_Boundary}
\begin{figure} 
\begin{center}
\includegraphics[scale=1.25]{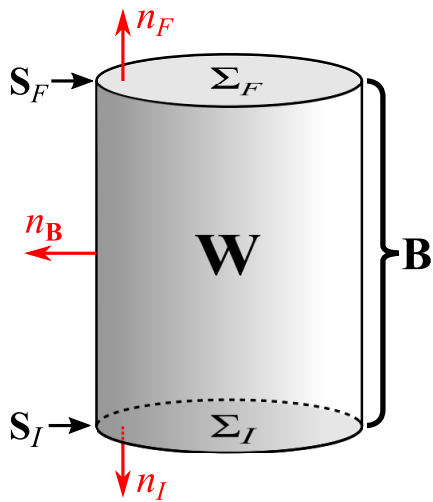}
\end{center}
\caption[Notation for Spacetime Boundary]{An illustration of a cylindrical boundary for a spacetime region \(\textbf{W}\), with boundary \(\partial \textbf{W}=\Sigma_I \cup \textbf{B} \cup \Sigma_F\). The vertical direction is timelike, so \(\Sigma_I\) and \(\Sigma_F\) are spacelike surfaces of codimension 1, and \(\textbf{B}\) is a timelike surface of codimension 1. The surfaces \(\textbf{S}_I\) and  \(\textbf{S}_F\) are surfaces of codimension 2 (assumed to have exclusively spacelike tangent vectors) that form boundaries between  \(\Sigma_I\), \(\textbf{B}\), and \(\Sigma_F\). The unit normal vectors (shown in red) are defined to be outward pointing; \(n_I=[n^\mu_I]\) is the unit normal to \(\Sigma_I\), \(n_{\textbf{B}}=[n^\mu_{\textbf{B}}]\) is the unit normal to \(\textbf{B}\), and \(n_F=[n^\mu_F]\) is the unit normal to \(\Sigma_F\). See Sec. \ref{Section:Gravitational Action_With_Spatial_Boundary}.} \label{3GR-BoundaryNotation}
\end{figure}

We now consider the case in which we consider spacetime regions \(\textbf{W}\) with a spatial boundary; in particular, we require \(\textbf{W}\) to have a boundary \(\partial \textbf{W}\) as described in Fig. \ref{3GR-BoundaryNotation}. The boundary \(\partial \textbf{W}\) consists of three regions, \(\Sigma_I\), \(\textbf{B}\) and \(\Sigma_F\), and is nonsmooth on the two-dimensional surfaces \(\textbf{S}_I\) and \(\textbf{S}_F\). We require the surfaces \(\Sigma_I\) and \(\Sigma_F\) to be spacelike, meaning that they admit a positive-definite induced metric \(\gamma_{ij}\) with signature \((+,+,+)\). On the other hand, we require the surface \(\textbf{B}\) to be timelike, meaning that the induced metric \(q_{ab}\) on \(\textbf{B}\) has a signature \((-,+,+)\).

This time, the gravitational action takes the form
\begin{equation} \label{3GR-GravitationalActionSpatialBoundaries}
\begin{aligned}
S_{GR,\textbf{B}}[g^{\mu \nu}]=\frac{1}{2 \kappa}\int_{\textbf{W}} \> R \>\sqrt{|g|} \> d^4 x -\frac{1}{\kappa}\int_{\Sigma_F} K \> \sqrt{|h|}  \> d^3 y+\frac{1}{\kappa}\int_{\textbf{B}} \underline{K} \> \sqrt{|q|}  \> d^3 y-\frac{1}{\kappa}\int_{\Sigma_I} K \> \sqrt{|h|}  \> d^3 y+S_C,
\end{aligned}
\end{equation}

\noindent where underlined quantities are defined on the boundary surface \(\textbf{B}\). For nonsmooth boundaries with spacelike junction surfaces \(\textbf{S}_I\) and \(\textbf{S}_F\), we must include the term \(S_C\), which is often referred to as the corner term \cite{Sorkin1975,*Sorkin1981,HartleSorkin1981,Hayward1993,BrillHayward1994} (also see Refs.  \cite{Parattu2016a,Parattu2016b,Lehner2016,Jubb2017,Chakraborty2017}),
\begin{equation} \label{3GR-HaywardTerm}
\begin{aligned}
S_C:=\frac{1}{\kappa}\int_{\textbf{S}_I} \eta_I \sqrt{|\sigma|}  \> d^2 z+\frac{1}{\kappa}\int_{\textbf{S}_F} \eta_F \sqrt{|\sigma|}  \> d^2 z,
\end{aligned}
\end{equation}

\noindent where the rapidity angles \(\eta_I\) and \(\eta_F\) are formed from the inner product between the unit normal vectors at the junction surfaces \(\textbf{S}_I\) and \(\textbf{S}_F\):
\begin{equation} \label{3GR-HaywardIntegrand}
\begin{aligned}
 &\eta_I:=\text{arcsinh}\left(\langle n_I,n_{\textbf{B}} \rangle|_{{\textbf{S}}_I}\right)\\
 &\eta_F:=\text{arcsinh}\left(\langle n_F,n_{\textbf{B}} \rangle|_{{\textbf{S}}_F}\right).
\end{aligned}
\end{equation}

\noindent We may ignore this term if the inverse metric tensor \(g^{\mu \nu}\) is held fixed at the junction surfaces \(\textbf{S}_I\) and \(\textbf{S}_F\), and the boundary displacements \(\delta x^\mu(y)\) vanish in a neighborhood of the junction surfaces \(\textbf{S}_I\) and \(\textbf{S}_F\). In other words, we require \(\langle n_I,n_{\textbf{B}} \rangle|_{{\textbf{S}}_I}\) and \(\langle n_F,n_{\textbf{B}} \rangle|_{{\textbf{S}}_F}\) to be held fixed and that \(\delta x^\mu|_{{\textbf{S}}_I}=0\), \(\delta x^\mu|_{{\textbf{S}}_F}=0\), \(\delta g_{\mu \nu}|_{{\textbf{S}}_I}=0\), and \(\delta g_{\mu \nu}|_{{\textbf{S}}_F}=0\).

Without loss of generality, we may choose the boundary displacement \(\delta x^\mu\) to take the following form: 
\begin{equation} \label{3GR-NormalBoundaryDisplacementSpatialPiecewise1}
\begin{aligned}
\delta x^\mu|_{\Sigma_I} &=n^\mu \, \Delta \tau_i(y) &\>\>\>\>\>\>\>\>\>\>&\text{for \(y \in {\Sigma_I}\)}\\
\delta x^\mu|_{\textbf{B}} &=n^\mu \, \Delta r(y) &\>\>\>\>\>\>\>\>\>\>&\text{for \(y \in \textbf{B}\)}\\
\delta x^\mu|_{\Sigma_F} &=n^\mu \, \Delta \tau_f(y) &\>\>\>\>\>\>\>\>\>\>&\text{for \(y \in {\Sigma_F}\)}.
\end{aligned}
\end{equation}

\noindent The Weiss variation of the action is (we stress that \(I\) and \(F\) are labels---they are \textit{not} indices to be summed over)
\begin{equation} \label{3GR-GravitationalActionVariationFullWeiss2b}
\begin{aligned}
\delta S_{GR,\textbf{B}}&=\frac{1}{2 \kappa}\int_{\textbf{W}} \, {G}_{\mu \nu} \, \delta g^{\mu \nu} \sqrt{|g|} \,  d^4 x + \int_{\Sigma_I} \biggl(P^I_{ij} \, \Delta h^{ij}_I  - \mathscr{H}_I \, \Delta \tau_I \biggr)d^3 y + \int_{ \textbf{B}} \biggl(\underline{P}_{ab} \> \Delta q^{ab}  - \mathscr{H}_{\textbf{B}} \, \Delta s \biggr) \, d^3 y\\
&\>\>\>\>\>   + \int_{\Sigma_F} \biggl(P^F_{ij} \> \Delta h^{ij}_F  - \mathscr{H}_F \, \Delta \tau_F \biggr) \, d^3 y
\end{aligned}
\end{equation}

\noindent where we define the momentum densities
\begin{equation} \label{3GR-MomentumDensitySpatial}
\begin{aligned}
{P}^I_{ij}&:=-\frac{1}{2 \kappa}\,({K}_{ij}-{K} \> h^I_{ij})  \, \sqrt{|h_I|}\\
\underline{P}_{ab}&:=\frac{1}{2 \kappa}\,(\underline{K}_{ab}-\underline{K} \> q_{ab})  \, \sqrt{|q|}\\
{P}^F_{ij}&:=-\frac{1}{2 \kappa}\,({K}_{ij}-{K} \> h^F_{ij})  \, \sqrt{|h_F|}
\end{aligned}
\end{equation}

\noindent and the Hamiltonian densities
\begin{equation} \label{3GR-HamiltonianDensitySpatial}
\begin{aligned}
\mathscr{H}_{I}&:=\frac{1}{2 \kappa} \, P_{ij}^I \, G^{ijkl} \, P_{kl}^I - \frac{1}{2\kappa} \,{^3}{R} \, \sqrt{|h_I|}\\
\mathscr{H}_{\textbf{B}}&:=-\frac{1}{2 \kappa} \, \underline{P}_{ab} \, \underline{G}^{abcd} \, \underline{P}_{cd} - \frac{1}{2\kappa} \,{^3}{R} \, \sqrt{|q|}\\
\mathscr{H}_{F}&:=\frac{1}{2 \kappa} \, P_{ij}^I \, G^{ijkl} \, P_{kl}^I - \frac{1}{2\kappa} \,{^3}{R} \, \sqrt{|h_F|},
\end{aligned}
\end{equation}

\noindent where \(\mathscr{H}_{I}\) is defined over \(\Sigma_I\) and \(\mathscr{H}_{F}\) is defined over \(\Sigma_F\) and the quantities \(G^{ijkl}\) and \(\underline{G}^{abcd}\) are given by
\begin{equation} \label{3GR-DeWittSupermetricSpatial}
\begin{aligned}
G^{ijkl}:=\frac{2 \, \kappa^2}{\sqrt{h}} \left ( h^{ik} \, h^{jl} + h^{il} \, h^{jk} - h^{ij} \, h^{kl}\right)\\
\underline{G}^{abcd}:=\frac{2 \, \kappa^2}{\sqrt{q}} \left (q^{ac} \, q^{bd} + q^{ad} \, q^{bc} - q^{ab} \, q^{cd} \right).
\end{aligned}
\end{equation}

\noindent Note that for vacuum solutions of the Einstein field equations, the right-hand sides of all three equations in (\ref{3GR-HamiltonianDensitySpatial}) vanish. We also note that the inverse 3-metric \(q^{ab}\) defines an extended superspace that is different than \(\mathscr{S}_{ex}(\Sigma)\) defined earlier, since \(q^{ab}\) defines a pseudo-Riemannian 3-geometry, rather than a Riemannian 3-geometry. We shall denote the extended superspace for pseudo-Riemannian 3-geometries on a 3-manifold \(\Sigma\) by \(\underline{\mathscr{S}}_{ex}(\Sigma)\). Underlined quantities will either refer to quantities defined on \(\underline{\mathscr{S}}_{ex}(\Sigma)\) or quantities defined on a pseudo-Riemannian 3-manifold. 

\section{From the Path Integral to the Schr{\"o}dinger Equation}
\subsection{Variation of the transition amplitude}
We now show that wave functions constructed from Feynman path integrals formally satisfy the Schr{\"o}dinger equation \cite{FeynmanHibbsQM}.\footnote{Our derivation at first follows closely the one found in the lecture notes \cite{BlauPILec,*MurayamaPILec}, but we extend it to include a discussion of the operator-path integral correspondence.} We begin by considering the position-basis transition amplitude
\begin{equation} \label{TransitionAmplitudePositionBasis}
\langle q^i_2,t_2|q^i_1,t_1 \rangle=\mathcal{K}( q^i_2,t_2;q^i_1,t_1),
\end{equation}

\noindent where \(|q^i,t\rangle\) is the position-basis state vector and the function \(\mathcal{K}( q^i_2,t_2;q^i_1,t_1)\) is given by the path integral expression [we shall henceforth refer to the function \(\mathcal{K}( q^i_2,t_2;q^i_1,t_1)\) as the transition amplitude]
\begin{equation} \label{4QM-TransitionAmplitudePathIntegral2}
\mathcal{K}( q^i_2,t_2;q^i_1,t_1) =\int \mathscr{D}q \> e^{(i/\hbar) \> S[q]},
\end{equation}

\noindent where the action functional \(S[q]\) (\ref{3M-Action}) is implicitly a function of the end point values \(q^i_1\), \(t_1\), \(q^i_2\), and \(t_2\) and \(\int \mathscr{D} {q}\) represents a measure\footnote{We absorb normalization factors into the definition of the measure.} on the space of functions \(q^i(t)\) subject to the Dirichlet boundary conditions \(q^i(t_1)=q^i_1\) and \(q^i(t_2)=q^i_2\); in particular, we hold the function \(q^i(t)\) fixed to the values \(q^i_1\) and \(q^i_2\) at the respective times \(t_1\) and \(t_2\). We do not provide a rigorous definition\footnote{For a rigorous treatment of path integrals, we refer the reader to the formalism of Cartier and DeWitt-Morette: Refs. \cite{CartierDeWitt-Morette2010,CartierDeWitt-Morette1995}.} for the measure; to proceed, we only need the property that the measure is invariant under shifts in the function \(q^i(t)\) of the form \(q^{i \, \prime} (t)=q^i(t)+\delta q^i(t)\). Under this shift, we require the measure to satisfy
\begin{equation} \label{4QM-MeasureInvarianceShift}
\int \mathscr{D} q^{\prime} \> F[q^\prime]= \int \mathscr{D} q \> F[q+\delta q]
\end{equation}

\noindent for some functional \(F[q]\). The property (\ref{4QM-MeasureInvarianceShift}) is motivated by the analogous property for the usual single-variable integral for shifts \(x^{\prime}=x+\delta x\) in the integration variable \(x\):
\begin{equation} \label{4QM-StdIntegralInvarianceShift}
\int_{-\infty}^{\infty} f(x^\prime) \, dx^{\prime} = \int_{-\infty}^{\infty} f(x+\delta x) \, dx.
\end{equation}

We now consider what happens when we vary the function \(q^i(t)\) while holding the end point values \(q^i_1\), \(t_1\), \(q^i_2\), and \(t_2\) fixed. We may use (\ref{4QM-MeasureInvarianceShift}) to write the following:
\begin{equation} \label{4QM-TransitionAmplitudePathIntegralShiftInvariance}\begin{aligned}
\int \mathscr{D}q^{ \prime} \> e^{(i/\hbar) \> S[q^{\prime}]}&=\int \mathscr{D}q \> e^{(i/\hbar) \> S[q+\delta q]}.
\end{aligned}
\end{equation}

\noindent We may perform a relabeling of the integration variable \(q^{i \> \prime} \rightarrow q^i\) to write
\begin{equation} \label{4QM-TransitionAmplitudePathIntegralChangeVariable}
\begin{aligned}
\int \mathscr{D}q^{ \prime} \> e^{(i/\hbar) \> S[q^{\prime}]}&=\int \mathscr{D}q \> e^{(i/\hbar) \> S[q]}
\end{aligned}
\end{equation}

\noindent Equations (\ref{4QM-TransitionAmplitudePathIntegralShiftInvariance}) and (\ref{4QM-TransitionAmplitudePathIntegralChangeVariable}) may be used to obtain:
\begin{equation} \label{4QM-TransitionAmplitudePathIntegralShift}
\int \mathscr{D}q \> e^{(i/\hbar) \> S[q+\delta q]}=\int \mathscr{D}q^i \> e^{(i/\hbar) \> S[q]} \>\>\>\>\>\>\>\> \Rightarrow \>\>\>\>\>\>\>\> \frac{i}{\hbar}\int \mathscr{D}q \> \delta_0 S \> e^{(i/\hbar) \> S[q]}=0
\end{equation}

\noindent where the subscript \(0\) in the variation of the action \(\delta_0 S\) indicates that the end point values \(q^i_1\), \(t_1\), \(q^i_2\), and \(t_2\) are held fixed. Note that, since \(\delta q^i(t)\) is independent of \(q^i(t)\), we may pull it out of the path integral. The demand that Eq. (\ref{4QM-TransitionAmplitudePathIntegralShift}) must hold for \textit{all} \(\delta q^i(t)\) implies
\begin{equation} \label{4QM-PathIntegralEhrenfestTheorem}
\int \mathscr{D}q \> \left(\frac{\partial L}{\partial q^i} - \frac{d}{dt} \left(\frac{\partial L}{\partial \dot{q}^i} \right) \right) \> e^{(i/\hbar) \> S[q]}=0.
\end{equation}

We now consider what happens when the end points are displaced; in particular, we change the values of \(q^i_1\), \(t_1\), \(q^i_2\), and \(t_2\). Equation (\ref{4QM-TransitionAmplitudePathIntegralChangeVariable}) is no longer valid, since the displacement of the end points prevents us from performing a change in the integration variable. The change in the transition amplitude is given by the following difference in path integrals:
\begin{equation} \label{4QM-VariationTransitionAmplitude}
\begin{aligned}
\delta \mathcal{K} & =\int \mathscr{D}q \> e^{(i/\hbar) \> S^\prime[q+\delta q]}-\int \mathscr{D}q \> e^{(i/\hbar) \> S[q]}.
\end{aligned}
\end{equation}

\noindent Since the measure itself is invariant under displacements of the end points (the displacements in \(t_1\) and \(t_2\) may be absorbed into a redefinition of \(t\)), the change in the amplitude is given by the change in the integrand. In particular, the change in the amplitude is given by the change in the action:
\begin{equation} \label{4QM-VariationTransitionAmplitude2}
\begin{aligned}
\delta \mathcal{K} & = \frac{i}{\hbar} \int \mathscr{D}q \> \delta S\> e^{(i/\hbar) \> S[q]}.
\end{aligned}
\end{equation}

\noindent Recalling the Weiss variational principle, we begin by performing the following infinitesimal transformation on the paths [recall (\ref{3M-WeissPaths}) and (\ref{3M-WeissTimeParam})]:
\begin{equation} \label{4QM-WeissTransformation}
\begin{aligned}
{q^{i \> \prime}}(t)&=q^{i}(t)+ \delta q^{i}(t)\\
{t^\prime}_1 &=t_1+ \Delta t_1\\
{t^\prime}_2 &=t_2+ \Delta t_2,
\end{aligned}
\end{equation}

\noindent where \(\delta q^{i}(t) = \varepsilon \> \eta^{i}(t)\) and we assume \(\Delta t_1 \propto \varepsilon\) and \(\Delta t_2 \propto \varepsilon\). If the action contains no more than first-order time derivatives in \(q^i\) (and a Legendre transformation can be performed), then to first order in \(\varepsilon\), the change in the action is given by the Weiss variation of the action (\ref{3M-WeissVariation}),
\begin{equation} \label{4QM-WeissVariation}
\begin{aligned}
\delta S &=\int^{t_2}_{t_1} \left(\frac{\partial L}{\partial q^{i}} - \frac{d}{dt} \left(\frac{\partial L}{\partial \dot{q}^{i}} \right)\right) \delta q^{i} \> dt + \left(p_i \> \Delta q^{i} - H \> \Delta t\right) \biggr |^{t_2}_{t_1},
\end{aligned}
\end{equation}

\noindent where \(p_i= {\partial L}/{\partial \dot{q}^{i}}\), \(H\) is the Hamiltonian, and
\begin{equation} \label{3M-EndpointDisplacementB}
\begin{aligned}
\Delta q^{i}_1&:={q^{i \>\prime}}({t^\prime}_1)-q^{i}(t_1)=\varepsilon (\eta^{i}(t_1) + \dot{q}^{i}(t_1))+O(\varepsilon^2)\\
\Delta q^{i}_2&:={q^{i \>\prime}}({t^\prime}_2)-q^{i}(t_2)=\varepsilon (\eta^{i}(t_2) + \dot{q}^{i}(t_2))+O(\varepsilon^2).
\end{aligned}
\end{equation}

\noindent Using (\ref{4QM-PathIntegralEhrenfestTheorem}), the change in the amplitude is given by
\begin{equation} \label{4QM-VariationTransitionAmplitude3}
\begin{aligned}
\delta \mathcal{K} & = \frac{i}{\hbar}\int \mathscr{D}q^i \>\left(p_i \> \Delta q^i - H \> \Delta t\right) \biggr |^{t_2}_{t_1} \> e^{(i/\hbar) \> S[q]}.
\end{aligned}
\end{equation}

\noindent Now, consider what happens when we hold \(q^i_1\) and \(t_1\) fixed, so that \(\Delta q^{A}_1=0\) and \(\Delta t_1=0\). We may write the above (\ref{4QM-VariationTransitionAmplitude3}) as
\begin{equation} \label{4QM-VariationTransitionAmplitude3b}
\begin{aligned}
\delta \mathcal{K} |_{(q^i_1,t_1) \rightarrow  \text{fixed}}& = \frac{i}{\hbar}\left(\int \mathscr{D}q \>p_{i,2} \> e^{(i/\hbar) \> S[q]} \right)  \Delta q^{i}_2\\
&\>\>\>\>\>-  \frac{i}{\hbar}\left(\int \mathscr{D}q \>H|_{t_2} \> e^{(i/\hbar) \> S[q]}\right) \Delta t_2,
\end{aligned}
\end{equation}

\noindent where we introduce the notation \(p_{i,1}=p_i |_{t_1}\) and \(p_{i,2}=p_i |_{t_2}\). Given an initial state function \(\psi(q^i_1)\) (equivalent to the state vector \(|q^i,t_1\rangle\)), where \(\psi(\cdot)\) is some normalizable complex function, we may define the wave function \(\Psi(q^i_2,t_2)\) in terms of the transition amplitude in the following way:
\begin{equation} \label{4QM-TransitionAmplitudeWavefunction}
\begin{aligned}
\Psi(q^i_2,t_2) := \int \mathcal{K}( q^i_2,t_2;q^i_1,t_1) \, \psi(q^i_1) \, d q^i_1.
\end{aligned}
\end{equation}

\noindent To simplify the analysis, we consider the case in which the initial state \(\psi(q^i_1)\) is sharply peaked around some value for \(q^i\); we may recover the general results by integrating the results over all values of \(q^i\) [in particular, for some path integral expression \(PI=PI( q^i_2,t_2;q^i_1,t_1)\), we perform the integral \(\int PI \, \psi(q^i_1) \, d q^i_1\)].

One may infer from Eq. (\ref{4QM-VariationTransitionAmplitude3b}) the expression for the differential of \(\Psi(q^i_2,t_2)\),
\begin{equation} \label{4QM-WavefunctionDifferential}
\begin{aligned}
d \Psi& = \frac{i}{\hbar} \left(\int \mathscr{D}q \>p_{i,2} \> e^{(i/\hbar) \> S[q]} \right)  d q^i_2-  \frac{i}{\hbar} \left(\int \mathscr{D}q \>H|_{t_2} \> e^{(i/\hbar) \> S[q]}\right) d t_2,
\end{aligned}
\end{equation}

\noindent which in turn yields the following expressions for the derivatives of the wave function:
\begin{equation} \label{4QM-WavefunctionHamiltonianOperatorPI}
\begin{aligned}
&\frac{i}{\hbar} \> \int \mathscr{D}q \>H|_{t_2} \> e^{(i/\hbar) \> S[q]}=-\frac{\partial \Psi}{\partial t_2}\\
\end{aligned}
\end{equation}
\begin{equation} \label{4QM-WavefunctionMomentumOperatorPI}
\begin{aligned}
&\frac{i}{\hbar} \>\int \mathscr{D}q \>p_{i,2} \> e^{(i/\hbar) \> S[q]}=\frac{\partial \Psi}{\partial q^{A}_2}.
\end{aligned}
\end{equation}

\noindent Equation (\ref{4QM-WavefunctionMomentumOperatorPI}) motivates the following definition for the momentum operator for wave functions \(\Psi(q_2,t_2)\) written in the position basis:
\begin{equation} \label{4QM-MomentumOperator}
\begin{aligned}
& \hat{p}_{i,2} := -i \> \hbar \>\frac{\partial}{\partial q^{i}_2}.
\end{aligned}
\end{equation}

We may motivate the Schr{\"o}dinger equation from (\ref{4QM-WavefunctionHamiltonianOperatorPI}), but to obtain an unambiguous expression for the  Schr{\"o}dinger equation, we must first establish an operator-path integral correspondence. In particular, we must construct an operator corresponding to the path integral:
\begin{equation} \label{4QM-pqsInPI}
\int \mathscr{D}q \>(p_{i_1}...p_{i_n} \> q^{j_1}...  q^{j_m})|_{t_2} \>  \> e^{(i/\hbar) \> S[q]}
\end{equation}

\noindent This is often referred to as the problem of operator ordering. In the next section, we shall explore this further.  If the Hamiltonian \(H|_{t_2}\) can be expressed as a polynomial function of \(p_{i,2}\) and \(q^i_2\), the path integral (\ref{4QM-WavefunctionHamiltonianOperatorPI}) consists of terms of the form (\ref{4QM-pqsInPI}). If one can establish a correspondence rule for operators and path integrals of the form (\ref{4QM-pqsInPI}), one may construct a Hamiltonian operator \(\hat{H}|_{t_2}=\hat{H}(\hat{p}_i,\hat{q}^i)|_{t_2}\); Eq. (\ref{4QM-WavefunctionHamiltonianOperatorPI}) may be rewritten as
\begin{equation} \label{4QM-SchrodingerEquationPI}
\begin{aligned}
&i \> \hbar \> \frac{\partial \Psi}{\partial t_2}=\hat{H}(\hat{p}_i,\hat{q}^i)|_{t_2} \, \Psi,
\end{aligned}
\end{equation}

\noindent which is, of course, the time-dependent Schr{\"o}dinger equation.

\subsection{Path integral-operator correspondence rule}
We now motivate a correspondence rule for path integral expressions and their operator counterparts. We begin by establishing an operator expression for factors of \(q^i\) that appear in the path integral. Since \(q^i_2\) is evaluated at the end point, we may trivially write
\begin{equation} \label{4QM-Wavefunctionq}
\begin{aligned}
&\int \mathscr{D}q \>q^i_2 \> e^{(i/\hbar) \> S[q]}=q^i_2 \int \mathscr{D}q \> e^{(i/\hbar) \> S[q]} =q^i_2 \> \Psi\\
&\int \mathscr{D}q \>q^{i_1}_2 ...q^{i_n}_2  \> e^{(i/\hbar) \> S[q]}=\left(q^{i_1}_2 ...q^{i_n}_2  \right)\int \mathscr{D}q \> e^{(i/\hbar) \> S[q]} =\left(q^{i_1}_2 ...q^{i_n}_2  \right) \> \Psi,
\end{aligned}
\end{equation}

\noindent which motivates the following definition for the operator \(\hat{q}^i_2\):
\begin{equation} \label{4QM-qoperator}
\begin{aligned}
\hat{q}^i_2 \> \Psi &= q^i_2 \> \Psi \\
\hat{q}^{i_1}_2 ...\hat{q}^{i_n}_2 \> \Psi &=\left(q^{i_1}_2 ...q^{i_n}_2  \right) \> \Psi.
\end{aligned}
\end{equation}

\noindent The definition (\ref{4QM-qoperator}) for the operator \(\hat{q}^i_2\) may be used to establish an operator expression for the following path integral:
\begin{equation} \label{4QM-Wavefunctionpq}
\begin{aligned}
&\int \mathscr{D}q \>(p_i \> q^j)|_{t_2} \>  \> e^{(i/\hbar) \> S[q]}.
\end{aligned}
\end{equation}

\noindent Naively, the above expression may be identified with the operation \((\hat{p}_i \> \hat{q}^j )|_{t_2} \>\Psi\) or \((\hat{q}^j \>\hat{p}_i)|_{t_2}\>\Psi\). The reader familiar with quantum mechanics will be fully aware of the fact that these two operations are inequivalent, since the operators \(\hat{p}_i|_{t_2}\) and \(\hat{q}^i_2\) do not commute. Indeed, we obtain the explicit expressions (\(\delta^j_i\) being the Kronecker delta)
\begin{equation} \label{4QM-Wavefunctionpqqp}
\begin{aligned}
&(\hat{p}_i \> \hat{q}^j)|_{t_2}  \>\Psi= -i \> \hbar \>\frac{\partial}{\partial q^{i}_2} (q^j_2 \>\Psi)= -i \> \hbar \left(\delta^j_i \> \Psi+ q^j_2\>\frac{\partial \Psi}{\partial q^{i}_2} \right)\\
& (\hat{q}^j \> \hat{p}_i)|_{t_2} \>\Psi= -i \> \hbar \> q^i_2 \>\frac{\partial \Psi}{\partial q^{i}_2},
\end{aligned}
\end{equation}

\noindent which yield the commutation relation when combined:
\begin{equation} \label{4QM-Commutator}
\begin{aligned}
&([\hat{p}_i , \hat{q}^j])|_{t_2}=(\hat{p}_i \> \hat{q}^j- \hat{q}^j \> \hat{p}_i)|_{t_2}= -i \> \hbar \> \delta^j_i .
\end{aligned}
\end{equation}

We now establish the operator expression corresponding to (\ref{4QM-Wavefunctionpq}). To do this, we expand the second line of (\ref{4QM-Wavefunctionpqqp}) in the manner
\begin{equation} \label{4QM-Wavefunctionqp}
\begin{aligned}
(\hat{q}^j \> \hat{p}_i)|_{t_2} \>\Psi&= -i \> \hbar \> q^j_2 \>\frac{\partial \Psi}{\partial q^{i}_2} =q^j_2 \int \mathscr{D}q \> p_{i,2} \> e^{(i/\hbar) \> S[q]} = \int \mathscr{D}  q \>q^j_2 \>p_{i,2} \> e^{(i/\hbar) \> S[q]}\\
& = \int \mathscr{D} q  \> (q^j \>p_i) |_{t_2} \> e^{(i/\hbar) \> S[q]}= \int \mathscr{D}  q \> (p_i \> q^j) |_{t_2} \> e^{(i/\hbar) \> S[q]},
\end{aligned}
\end{equation}

\noindent where the last equality follows from the expression\footnote{We assume that \(q^i\) and \(p_i\) are \textit{c}-number valued; the present analysis excludes Grassmann-valued degrees of freedom.} \(q^j \>p_i=p_i \> q^j\). The computation (\ref{4QM-Wavefunctionqp}) demonstrates that the path integral expression in (\ref{4QM-Wavefunctionpq}) corresponds to the operator ordering \((\hat{q}^B \> \hat{p}_i)|_{t_2} \>\Psi\), in which the momentum operator \(\hat{p}_{i,2}\) appears to the \textit{right}\footnote{By this, we mean that the momentum operator \(\hat{p}_{i,2}\) is applied to \(\Psi\) first.} of the position operator. The correspondence suggested by (\ref{4QM-Wavefunctionqp}) applies only to wave functions \(\Psi(q_2,t_2)\) in the \textit{position} basis; the correspondence rule for path integrals of the form (\ref{4QM-Wavefunctionpq}) is basis dependent.

One may expect that, given the definition (\ref{4QM-MomentumOperator}) for the momentum operator \(\hat{p}_i\), one has the following correspondence rule:
\begin{equation} \label{4QM-WavefunctionMomentumOperatorHigher}
\begin{aligned}
&\left( {i}/{\hbar}\right)^n \>\int \mathscr{D}q \>(p_{i_1} ... p_{i_n}) |_{t_2} \> e^{(i/\hbar) \> S[q]}\approx\left({i}/{\hbar}\right)^n \, (\hat{p}_{i_1} ... \hat{p}_{i_n}) |_{t_2} \, \Psi = \frac{\partial^n \Psi}{\partial q^{i_1}_2 ... \partial q^{i_n}_2}\\
\end{aligned}
\end{equation}

\noindent The difficulty with this correspondence rule (\ref{4QM-WavefunctionMomentumOperatorHigher}) is that it is not compatible with coordinate transformations on the configuration space \(\mathbb{Q}\). In particular, we expect \(\Psi\) and the Schr{\"o}dinger equation to transform as scalars under coordinate transformations on \(\mathbb{Q}\), but this is not in general true for a Schr{\"o}dinger equation constructed using the correspondence rule (\ref{4QM-WavefunctionMomentumOperatorHigher}). To see this, consider the expression:
\begin{equation} \label{4QM-ScalarOperatorMomFactors}
\begin{aligned}
&\left( {i}/{\hbar}\right)^n \>\int \mathscr{D}q \>(T^{i_1 ... i_n} \, p_{i_1} ... p_{i_n}) |_{t_2} \> e^{(i/\hbar) \> S[q]} \approx \left(T^{i_1 ... i_n}|_{t_2}\right)\frac{\partial^n \Psi}{\partial q^{i_1}_2 ... \partial q^{i_n}_2},
\end{aligned}
\end{equation}

\noindent where \(T^{i_1 ... i_n}=T^{i_1 ... i_n}(q)\) is a tensor on the configuration space manifold \(\mathbb{Q}\). Since the conjugate momenta \(p_i\) transform as cotangent vectors, the integrand on the left-hand side of (\ref{4QM-ScalarOperatorMomFactors}) transforms as a scalar (the action \(S[q]\) is assumed to be invariant under coordinate transformations). For \(n>1\), the right-hand side of (\ref{4QM-WavefunctionMomentumOperatorHigher}) does not transform as a scalar under coordinate transformations on \(\mathbb{Q}\).

The problems with establishing a correspondence rule may be attributed to ambiguities in the definition of the measure. Suppose that, for a given definition of the measure \(\int \mathscr{D} q\), one obtains the correspondence rule (\ref{4QM-WavefunctionMomentumOperatorHigher}). Since the integrand of (\ref{4QM-WavefunctionMomentumOperatorHigher}) transforms as a scalar, any nonscalar transformation law must come from the transformation of the measure \(\int \mathscr{D} q\). Thus, any nontensorial transformation law for the left-hand side of (\ref{4QM-WavefunctionMomentumOperatorHigher}) must come from the measure. Though we do not propose a definition for the measure, we require the measure to satisfy the property that the left-hand side of (\ref{4QM-ScalarOperatorMomFactors}) transforms as a scalar. This suggests that the measure yields the correspondence rule
\begin{equation} \label{4QM-CorrespondenceRule}
\begin{aligned}
\int \mathscr{D}q \>(p_{i_1} ... p_{i_n}) |_{t_2} \> e^{(i/\hbar) \> S[q]} &= (- i \, \hbar)^n  (\nabla_{i_1} ... \nabla_{i_n})|_{t_2} \, \Psi \\
\int \mathscr{D}q \>(p_{i_1}...p_{i_n} \> q^{j_1}...  q^{j_m})|_{t_2} \>  \> e^{(i/\hbar) \> S[q]} &= (\hat{q}^{j_1} ...  \hat{q}^{j_m}\,\nabla_{i_1} ... \nabla_{i_n})|_{t_2} \, \Psi ,
\end{aligned}
\end{equation}

\noindent where \(\nabla_i|_{t_2}\) is a connection (covariant derivative) on the configuration space manifold \(\mathbb{Q}\), constructed from the partial derivatives \(\partial /\partial q^i_2\) (with \(q^i_2 \in \mathbb{Q}\)) and connection coefficients \(\Gamma^i_{jk}\); for a second rank tensor \(T^i_j\), it takes the form
\begin{equation} \label{4QM-CovariantDeriv}
\begin{aligned}
\nabla_i  T^j_k |_{t_2}=\frac{\partial  T^i_j}{\partial q^i_2}+ \Gamma^j_{i l}\, T^l_k - \Gamma^l_{i k}\, T^j_l .
\end{aligned}
\end{equation}

\noindent The connection coefficients \(\Gamma^i_{jk}\) are assumed to be functions of \(q^i_2\) and are defined to satisfy the following transformation rule:
\begin{equation} \label{1-ConnectionCoeffTransformation} 
{\Gamma^\prime}^i_{jk}=\left(\frac{\partial {q^\prime}^i}{\partial q^a}\frac{\partial q^b}{\partial {q^\prime}^j}\frac{\partial q^c}{\partial {q^\prime}^k}\right){\Gamma}^a_{bc}-\frac{\partial q^b}{\partial {q^\prime}^j}\frac{\partial q^a}{\partial {q^\prime}^k}\left(\frac{\partial^2 {q^\prime}^i}{\partial
q^b \, \partial q^a}\right) .
\end{equation}

\noindent For a particle on a Riemannian manifold, it is natural to choose \(\Gamma^i_{jk}\) to be the Christoffel symbols. 

We stress that the correspondence rule (\ref{4QM-CorrespondenceRule}) for factors of the momenta is a \textit{property} that we require of the measure, rather than a derived result. An important question that should be addressed is whether one can explicitly construct a time-sliced path integral that yields the correspondence rule (\ref{4QM-CorrespondenceRule}). For our purposes, the full correspondence rule (\ref{4QM-CorrespondenceRule}) is not necessary, as we only need a correspondence rule for the following expression:
\begin{equation} \label{4QM-OperatorLaplaceBeltrami}
\begin{aligned}
\int \mathscr{D}q \, (g^{ij} \, p_i \, p_j)|_{t_2} \> e^{ \frac{i}{\hbar} \, S[q]}=-\hbar^2 g^{ij} \nabla_i  \nabla_j \Psi=- \hbar^2 \frac{1}{\sqrt{g}} \frac{\partial}{\partial q^i_2} \left(\sqrt{g} \, g^{ij} \,   \frac{\partial \Psi}{\partial q^j_2} \right) .
\end{aligned}
\end{equation}

\noindent This is, of course, the well-known Laplace-Beltrami ordering rule \cite{DeWitt1967,ChaichianDemichevPI,RyanTurbiner2004}. From the coordinate-invariant time-sliced definitions of the path integral presented in Ref. \cite{DeWitt1957} (also see Ref. \cite{ChaichianDemichevPI}), one may infer that the time-sliced measure yields the Laplace-Beltrami correspondence
\begin{equation} \label{4QM-FunctionalMeasureDefinitionInvLimit}
\int \mathcal{D} {q} \> (\cdot) :=\lim_{n \rightarrow \infty} \> \prod^{n-1}_{k=2}  \int_{\mathbb{Q}} d^Nq_{\tau_k} \> \sqrt{|M|} \exp \left(\frac{i \, \hbar}{6} \int R \, dt\right) (\cdot) ,
\end{equation}

\noindent where we use the replacement \(\dot{q}^i_{\tau_k}\rightarrow(q^i_{\tau_{k}}-q^i_{\tau_{k-1}})/\Delta t\) in the argument of the right-hand side. Though we note that the Laplace-Beltrami ordering (\ref{4QM-OperatorLaplaceBeltrami}) follows from our correspondence rule (\ref{4QM-CorrespondenceRule}), it is not yet clear to us that the time-sliced measure (\ref{4QM-FunctionalMeasureDefinitionInvLimit}) yields the correspondence rule (\ref{4QM-CorrespondenceRule}). We will not investigate the compatibility of the measure (\ref{4QM-FunctionalMeasureDefinitionInvLimit}) and the correspondence rule (\ref{4QM-CorrespondenceRule}) and leave it for future work---this question is beyond the scope of this article, as the Laplace-Beltrami ordering is sufficient for our purposes. In the next section, we use a Laplace-Beltrami-type ordering for the Wheeler-DeWitt equation by demanding that transition amplitudes be invariant under coordinate transformations on superspace.

\section{Wheeler-DeWitt Equation}

Again, we consider a spacetime \(\mathcal{M}\) that has no spatial boundary, so that it has the topology \(\mathbb{R} \times \Sigma^3\), where \(\Sigma^3\) is a three-dimensional manifold without a boundary. We begin with the gravitational action \(S_{GR}\), (\ref{3GR-GravitationalAction}), which we rewrite here,
\begin{equation} \label{4QG-GravitationalAction}
\begin{aligned}
S_{GR}[g^{\mu \nu}]&:=S_{EH}[g^{\mu \nu}]+S_{GHY}\;=\frac{1}{2 \kappa}\int_{\textbf{U}} \> R \>\sqrt{|g|} \> d^4 x+\frac{1}{\kappa}\int_{\partial \textbf{U}} K \> \varepsilon \> \sqrt{|\gamma|}  \> d^3 y ,
\end{aligned}
\end{equation}

\noindent where \(\textbf{U}\) has no spatial boundary (or where \(\partial \textbf{U}=\Sigma_I \cup \Sigma_F\), where \(\Sigma_I\) and \(\Sigma_F\) are spacelike hypersurfaces without a boundary). The most straightforward approach to the quantization of the gravitational field is to simply write down the path integral\footnote{It is well known that quantum general relativity [as given by functional integrals of the form (\ref{4QG-GravitationalPathIntegral})] is perturbatively nonrenormalizable (see Refs. \cite{Woodard2009} and \cite{Kiefer2012QG} for an overview). The analysis we present is formally nonperturbative, so the question of perturbative renormalizability will not enter into our analysis [one might imagine that in writing down the path integral (\ref{4QG-GravitationalPathIntegral}) we are studying the low-energy limit of some effective field theory for gravity].} (first written down in Ref. \cite{Misner1957})
\begin{equation} \label{4QG-GravitationalPathIntegral}
\begin{aligned}
\mathcal{K} \llbracket \gamma^{ij}_{F}; \gamma^{ij}_{I} \rrbracket=\int \mathscr{D} [g^{\alpha \beta}] \, e^{\frac{i}{\hbar} S_{GR}[g^{\alpha \beta}]},
\end{aligned}
\end{equation}

\noindent where the brackets \(\llbracket \, \rrbracket\) indicate a functional of functions defined over the boundary surface \(\partial \textbf{U}\) and \(\int \mathscr{D} [g^{\alpha \beta}]\) denotes a functional integral over all functions \(g^{\alpha \beta}(x)\) up to those that differ by a diffeomorphism. We require the integration measure \(\int \mathscr{D} [g^{\alpha \beta}]\) to be invariant under the shift (field redefinition) \(g^{\mu \nu}\rightarrow g^{\mu \nu}+ \delta g^{\mu \nu}\),\footnote{For the present analysis, we ignore the details of gauge fixing for the measure \(\int \mathscr{D} [g^{\alpha \beta}]\) or any other procedure for modding out diffeomorphisms; we only assume that the shift \(\delta g^{\mu \nu}\) is compatible with the procedure for gauge fixing or modding out diffeomorphisms.}, and we also require the measure to be defined so that the resulting amplitude \(\mathcal{K} \llbracket \gamma^{ij}_{F}; \gamma^{ij}_{I} \rrbracket\) is independent of coordinate transformations on \(\Sigma_I\) and \(\Sigma_F\). The path integral (\ref{4QG-GravitationalPathIntegral}) formally defines an unnormalized transition amplitude between a spacetime with an initial inverse 3-metric \(\gamma^{ij}_{I}:=\gamma^{ij}|_{\Sigma_I}\) for the initial hypersurface \(\Sigma_I\) and a final inverse 3-metric \(\gamma^{ij}_{F}:=\gamma^{ij}|_{\Sigma_F}\) for the hypersurface \(\Sigma_F\).

In this section we attempt to derive a Schr{\"o}dinger equation from the transition amplitude (\ref{4QG-GravitationalPathIntegral}), but we instead find that the transition amplitude must lie in the kernel of the formal Hamiltonian operator; the transition amplitude is independent of the time parameters \(t_I\) and \(t_F\).  This is known as the \textit{problem of time}\index{Quantum Gravity!Problem of Time}, and though it is primarily discussed in the context of the canonical formulation of quantum general relativity, we shall show in the next section that it is also present in the path integral formulation.

\subsection{Variation of the path integral}
We now perform the variation of the path integral (\ref{4QG-GravitationalPathIntegral}), assuming that the measure \(\int \mathscr{D} [g^{\alpha \beta}]\) is invariant under the shift \(g^{\mu \nu}\rightarrow g^{\mu \nu}+ \delta g^{\mu \nu}\). The variation of the path integral may then be written in terms of the variation of the action:
\begin{equation} \label{4QG-VariationGravitationalPathIntegral}
\begin{aligned}
\delta \mathcal{K} =\frac{i}{\hbar} \int \mathscr{D} [g^{\alpha \beta}] \, \delta S_{GR}\, e^{\frac{i}{\hbar} S_{GR}[g^{\alpha \beta}]}.
\end{aligned}
\end{equation}

\noindent Earlier, we presented the Weiss variation of the gravitational action (\ref{3GR-GravitationalActionVariationFullWeiss}),
\begin{equation} \label{4QG-GravitationalActionVariationFullWeiss}
\begin{aligned}
\delta S_{GR}&=\frac{1}{2 \kappa}\int_{\textbf{U}} \left({R}_{\mu \nu} -  \frac{1}{2} \>R \> g_{\mu \nu} \right) \delta g^{\mu \nu} \sqrt{|g|} \>  d^4 x+ \frac{ \varepsilon}{2\kappa}\int_{\partial \textbf{U}} \biggl(p_{\mu \nu} \> \Delta \gamma^{\mu \nu} \\
&\>\>\>\>\> +\left[ 2 \> D_\alpha p^{\alpha \beta} \> \gamma_{\mu \beta} +  n_\mu \left({^3}{R}-\varepsilon(K^2-K_{ij} \> K^{ij})\right)\right]\delta x^\mu \biggr) \> \sqrt{|\gamma|} \> d^3 y,
\end{aligned}
\end{equation}

\noindent where the boundary of \(\textbf{U}\) is given by the expression \(\partial \textbf{U}=\Sigma_I \cup \Sigma_F\), and [cf. Eq. (\ref{3GR-ConjugateMomentumTensor})]:
\begin{equation} \label{4QG-ConjugateMomentum}
\begin{aligned}
p_{\mu \nu}&:=(K_{\mu \nu} - K \> \gamma_{\mu \nu}).
\end{aligned}
\end{equation}

\noindent Note that the indices of \(p_{\mu \nu}\) are tangent to the boundary \(\partial \textbf{U}\). From the invariance of the path integral (\ref{4QG-VariationGravitationalPathIntegral}) under a change in integration variable [recall Eq. (\ref{4QM-TransitionAmplitudePathIntegralShiftInvariance})] and the invariance of the measure \(\int \mathscr{D} [g^{\alpha \beta}]\) under shifts of the form \(g^{\mu \nu}\rightarrow g^{\mu \nu}+ \delta g^{\mu \nu}\), one may show the following:
\begin{equation} \label{4QG-VariationGravitationalPathIntegralBulk}
\begin{aligned}
&\int \mathscr{D} [g^{\alpha \beta}] \,\int_{\textbf{U}} \left({R}_{\mu \nu} -  \frac{1}{2} \>R \> g_{\mu \nu} \right) \delta g^{\mu \nu} \sqrt{|g|} \,  d^4 x\, e^{\frac{i}{\hbar} S_{GR}[g^{\alpha \beta}]}=0\\
&\Rightarrow \>\>\>\>\> \int_{\textbf{U}}\left(\int \mathscr{D} [g^{\alpha \beta}] \, \left({R}_{\mu \nu} -  \frac{1}{2} \>R \> g_{\mu \nu} \right) \sqrt{|g|}\, e^{\frac{i}{\hbar} S_{GR}[g^{\alpha \beta}]} \right) \delta g^{\mu \nu} \,  d^4 x=0\\
&\Rightarrow \>\>\>\>\>\int \mathscr{D} [g^{\alpha \beta}] \, \left({R}_{\mu \nu} -  \frac{1}{2} \>R \> g_{\mu \nu} \right) \sqrt{|g|}\, e^{\frac{i}{\hbar} S_{GR}[g^{\alpha \beta}]}=0.
\end{aligned}
\end{equation}

\noindent The second line comes from the fact that the shift function \(\delta g^{\mu \nu}\) can be chosen independently of the integration variable \(g^{\mu \nu}\), and the last line can be inferred from the second line by requiring the second line to hold for all choices of \(\delta g^{\mu \nu}\). The last line is the statement that in the absence of matter the vacuum Einstein field equations are satisfied within the path integral. The variation of the path integral thus becomes
\begin{equation} \label{4QG-VariationGravitationalPathIntegral2}
\begin{aligned}
\delta \mathcal{K} &= \frac{i}{\hbar} \int \mathscr{D} [g^{\alpha \beta}] \, \biggl[ \int_{\partial \textbf{U}} \biggl(P_{\mu \nu} \> \Delta \gamma^{\mu \nu} - \mathscr{H}^\nu{_\mu} \, n_\nu \, \delta x^\mu \biggr) d^3 y \biggr]e^{\frac{i}{\hbar} S_{GR}[g^{\alpha \beta}]},
\end{aligned}
\end{equation}

\noindent where we have defined the following:
\begin{equation} \label{4QG-MomentumDef}
\begin{aligned}
P_{\mu \nu }:=\frac{ \varepsilon}{2\kappa} \, p_{\mu \nu } \, \sqrt{|\gamma|}=\frac{ \varepsilon}{2\kappa} \, (K_{\mu \nu } - K \> \gamma_{\mu \nu }) \, \sqrt{|\gamma|}
\end{aligned}
\end{equation}
\begin{equation} \label{4QG-HamiltonianTensorDef}
\begin{aligned}
\mathscr{H}^\nu{_\mu}:=-\frac{ \varepsilon}{2\kappa} \, \biggl[ 2 \, \varepsilon \, D_\alpha p^{\alpha \beta} \, \gamma_{\mu \beta} \, n^\nu +  \delta^\nu_\mu \, \left({^3}{R}-\varepsilon(K^2-K_{ij} \> K^{ij})\right)\biggr] \sqrt{|\gamma|}.
\end{aligned}
\end{equation}

\noindent We choose \(\delta x^\mu\) to take the form
\begin{equation} \label{4QG-NormalBoundaryDisplacement2}
\begin{aligned}
\delta x^\mu =n^\mu \, \Delta t ,
\end{aligned}
\end{equation}

\noindent where \(\Delta t\) is a constant. We then make use of the expressions \(p_{\mu \nu} \> \Delta \gamma^{\mu \nu}=p_{ij} \> \Delta \gamma^{ij}\) and \(\gamma_{\mu \beta}\, n^\mu=0\) to obtain the variation of the gravitational path integral,
\begin{equation} \label{4QG-VariationGravitationalPathIntegral2b}
\begin{aligned}
\delta \mathcal{K} &= \frac{i}{\hbar} \int \mathscr{D} [g^{\alpha \beta}] \, \biggl[ \int_{\partial \textbf{U}} \biggl(P_{ij} \> \Delta \gamma^{ij} - \mathscr{H}_{gf} \, \Delta t \biggr) d^3 y \biggr]e^{\frac{i}{\hbar} S_{GR}[g^{\alpha \beta}]},
\end{aligned}
\end{equation}

\noindent where \(\mathscr{H}_{gf}\) is given by (\ref{3GR-HamiltonianDensityGaugeFixed}), which we rewrite here:
\begin{equation} \label{4QG-HamiltonianDensityGaugeFixed}
\begin{aligned}
\mathscr{H}_{gf}= - \frac{1}{2\kappa} \, \biggl[{^3}{R}+K^2-K_{ij} \> K^{ij} \biggr]\sqrt{|\gamma|}.
\end{aligned}
\end{equation}

\noindent In the above expression, we have set \(\varepsilon=-1\), appropriate for the case of spacelike boundary surfaces in general relativity.

\subsection{Wave functional, operators, and commutators}
We begin defining an initial state \(\psi \llbracket \gamma^{ij}_{I} \rrbracket\) for some value of \(t_I\). The \textit{wave functional} may be formally\footnote{Since it depends on the formally defined measure \(\int \mathscr{D} [\gamma^{kl}_I]\), the right-hand side of (\ref{4QG-WaveFunctional}) is only defined in a formal sense. This is essentially the problem of defining the inner product (see Sec. 5.2.2 of Ref. \cite{Kiefer2012QG}, and also see Ref. \cite{Woodard1993} for a gauge-fixed definition of the inner product measure). If superspace has countable dimension, it may be possible in the superspace representation to define an inner product in the usual manner---this will be left for future investigation.} defined as:
\begin{equation} \label{4QG-WaveFunctional}
\begin{aligned}
\Psi=\Psi  \llbracket \gamma^{ij} \rrbracket (t):= \int \mathscr{D} [\gamma^{kl}_I] \> \mathcal{K}\llbracket \gamma^{ij} ;\gamma^{kl}_{I} \rrbracket \> \psi \llbracket \gamma^{kl}_{I} \rrbracket,
\end{aligned}
\end{equation}

\noindent where we substitute \(t\) in place of \(t_F\) (the surface  \(\Sigma_F\) will be denoted \(\Sigma_t\)) and \(\int \mathscr{D} [\gamma^{kl}_I]\) denotes a functional integral over the inverse 3-metric \(\gamma^{ij}_{I}\) on the surface \(\Sigma_I\) (up to coordinate transformations on \(\Sigma_I\)).\footnote{Since we have required \(\mathcal{K} \llbracket \gamma^{ij}_{F}; \gamma^{ij}_{I} \rrbracket\) to be independent of coordinate transformations on \(\Sigma_I\) and \(\Sigma_F\), the wave functional \(\Psi  \llbracket \gamma^{ij} \rrbracket\) must be independent of coordinate transformations on \(\Sigma_F\).} As before (in the derivation of the quantum mechanical Schr{\"o}dinger equation), we consider an initial state \(\Psi \llbracket \gamma^{ij}_{I} \rrbracket\) that is sharply peaked around some particular function \(\gamma^{ij}_{I}\) to eliminate the functional integral \(\int \mathscr{D} [\gamma^{kl}_I]\) (to recover the general result, reintroduce the functional integral \(\int \mathscr{D} [\gamma^{kl}_I]\)). To first order, infinitesimal changes in the wave functional \(\Psi\) will take the form
\begin{equation} \label{4QG-WaveFunctionalVariation}
\begin{aligned}
\delta \Psi=\int_{\Sigma_{t}} \frac{\delta \Psi}{\delta_{\Sigma_t} \gamma^{ij} } \, \Delta \gamma^{ij} \, d^3 y  + \frac{\partial\Psi}{\partial t} \, \Delta t,
\end{aligned}
\end{equation}

\noindent where we use the notation \(\delta /\delta_\Sigma \varphi\) to denote the functional derivative with respect to a function \(\varphi(x)\) restricted to a hypersurface \(\Sigma\); for a hypersurface with the parametrization \(x^\mu(y)\) and the function \(\bar{\varphi}(y):=\varphi |_{\Sigma}=\varphi(x(y))\), this means
\begin{equation} \label{4QG-RestrictedFunctionalDerivative}
\begin{aligned}
&\frac{\delta}{\delta_{\Sigma} \varphi}:=\frac{\delta }{\delta \bar{\varphi}}.
\end{aligned}
\end{equation}

\noindent Upon comparing (\ref{4QG-RestrictedFunctionalDerivative}) with (\ref{4QG-VariationGravitationalPathIntegral2b}), we identify the following expressions:
\begin{equation} \label{4QG-VariationGravitationalPathIntegralMom}
\begin{aligned}
\frac{\delta \Psi}{\delta_{\Sigma_t} \gamma^{ij} }&= \frac{i}{\hbar} \int \mathscr{D} [g^{\alpha \beta}] \, P_{ij} |_{\Sigma_{t}}  \, e^{\frac{i}{\hbar} S_{GR}[g^{\alpha \beta}]}\\
\end{aligned}
\end{equation}
\begin{equation} \label{4QG-VariationGravitationalPathIntegralHam}
\begin{aligned}
\frac{\partial\Psi}{\partial t} &=- \frac{i}{\hbar} \int \mathscr{D} [g^{\alpha \beta}] \, \biggl( \int_{\Sigma_t} \mathscr{H}_{gf} \, d^3 y \biggr) e^{\frac{i}{\hbar} S_{GR}[g^{\alpha \beta}]}.
\end{aligned}
\end{equation}

\noindent Equation (\ref{4QG-VariationGravitationalPathIntegralMom}) suggests that the momentum operator \(\hat{P}_{ij}\) satisfies
\begin{equation} \label{4QG-MomentumOperator}
\begin{aligned}
\hat{P}_{ij} \Psi:=-{i}\,{\hbar}\frac{\delta \Psi}{\delta_{\Sigma_t} \gamma^{ij}}.
\end{aligned}
\end{equation}

\noindent On the other hand, Eq. (\ref{4QG-VariationGravitationalPathIntegralHam}) suggests that we may formally define the quantum theory by constructing a Hamiltonian operator \(\hat{H}_{GR}\) such that its action on \(\Psi\) is formally equivalent to the following path integral:
\begin{equation} \label{4QG-HamiltonianOperatorCond}
\begin{aligned}
\hat{H}_{GR} \, \Psi&=\int \mathscr{D} [g^{\alpha \beta}] \, \biggl( \int_{\Sigma_t} \mathscr{H}_{gf} \, d^3 y \biggr) e^{\frac{i}{\hbar} S_{GR}[g^{\alpha \beta}]}.
\end{aligned}
\end{equation}

\noindent Since the wave functional \(\Psi\) is a functional of quantities defined on \(\Sigma_{t}\) only, a valid Hamiltonian operator must be defined in terms of operators at the surface \({\Sigma_{t}}\). Otherwise, the time evolution may become nonlinear in the time derivatives, and one is no longer doing quantum physics. To construct a Hamiltonian operator, we must first identify the operators that can be defined for the wave functional \(\Psi\). In general, an operator \(\hat{\mathcal{O}}\) acting on \(\Psi\) must have a path integral expression of the form
\begin{equation} \label{4QG-OperatorFormPathIntegral}
\begin{aligned}
\hat{\mathcal{O}} \, \Psi&=\int \mathscr{D} [g^{\alpha \beta}] \>{\mathcal{O}}|_{\Sigma_{t}} e^{i \, S_{Pal}/\hbar} ,
\end{aligned}
\end{equation}

\noindent where \({\mathcal{O}}|_{\Sigma_{t_F}}\) is some quantity defined on \({\Sigma_{t_F}}\). We have already identified the momentum operator \(\hat{P}_{ij}\) in Eq. (\ref{4QG-MomentumOperator}). We may define the inverse metric operator as follows (we explicitly write the arguments of \(\Psi\) for clarity\footnote{This is to indicate that the nonoperator quantities that appear outside of the wave functional depend on the values of the quantities that appear in the arguments of the wave functional.}):
\begin{equation} \label{4QG-InverseMetricOperator}
\begin{aligned}
\hat{\gamma}^{ij} \, \Psi  \llbracket \gamma^{ij} \rrbracket (t) :=\gamma^{ij}|_t \> \Psi  \llbracket \gamma^{ij} \rrbracket (t)=\int \mathscr{D} [g^{\alpha \beta}] \, \gamma^{ij}|_t \, e^{\frac{i}{\hbar} S_{GR}[g^{\alpha \beta}]} .
\end{aligned}
\end{equation}

\noindent The second equality is justified by the fact that, since \(\gamma^{ij}\) is held fixed at \(\Sigma_{t}\) in the path integral, we can pull factors of \(\gamma^{ij}|_{{t}}\) into the path integral, since \(\gamma^{ij}|_{{t}}\) is effectively constant with respect to the functional integral \(\int \mathscr{D} [g^{\alpha \beta}]\).

Since the 3-metric \(\gamma_{ij}\) and the volume element\footnote{We drop the absolute value symbols since we now assume \(\varepsilon=-1\).} \({\sqrt{\gamma}}\) can in principle be obtained algebraically from \(\gamma^{ij}\), then we can define a metric operator \(\hat{\gamma}_{ij}\) from \(\hat{\gamma}^{ij}\), as well as the operators \(\hat{\sqrt{\gamma}}\) and \((1/\hat{\sqrt{\gamma}})\) in the following manner (we explicitly write the arguments of \(\Psi\) for clarity):
\begin{equation} \label{4QG-MetricOperator}
\begin{aligned}
&\hat{\gamma}_{ij} \, \Psi  \llbracket \gamma^{ij} \rrbracket (t) :=\gamma_{ij}|_t \> \Psi  \llbracket \gamma^{ij} \rrbracket (t)\\
&\hat{\sqrt{\gamma}} \, \Psi  \llbracket \gamma^{ij} \rrbracket (t) :={\sqrt{\gamma}}|_t \> \Psi  \llbracket \gamma^{ij} \rrbracket (t)\\
&(1/\hat{\sqrt{\gamma}}) \, \Psi  \llbracket \gamma^{ij} \rrbracket (t) :=(1/{\sqrt{\gamma}})|_t \> \Psi  \llbracket \gamma^{ij} \rrbracket (t).
\end{aligned}
\end{equation}

\noindent Since the three-dimensional Ricci scalar \(^{(3)}R\) depends only on the values of the metric at the surface \(\Sigma_F\), we may construct the following operator for the Ricci scalar (we explicitly write the arguments of \(\Psi\) for clarity):
\begin{equation} \label{4QG-Ricci3CurvatureOperator}
\begin{aligned}
^{(3)}\hat{R} \, \Psi  \llbracket \gamma^{ij} \rrbracket (t) :={^{(3)}}R |_{\gamma_{ij}(t), \,t} \> \Psi  \llbracket \gamma^{ij} \rrbracket (t).
\end{aligned}
\end{equation}

To obtain commutation relations for the operators \(\hat{\gamma}^{ij}\) and \(\hat{P}_{ij}\), we first examine the expressions [cf. (\ref{4QM-Wavefunctionpqqp})]:
\begin{equation} \label{4QG-Wavefunctionpqqp}
\begin{aligned}
&(\hat{P}_{kl} \> \hat{\gamma}^{ij})|_{t}  \>\Psi= -i \> \hbar \>\frac{\delta}{\delta_{\Sigma_t} \gamma^{kl}} (\gamma^{ij}_F \>\Psi)= -i \> \hbar \left(\delta^i_k \, \delta^j_l \, \delta(y-y^\prime) \> \Psi+ \gamma^{ij}\>\frac{\delta \Psi}{\delta_{\Sigma_t} \gamma^{kl}} \right)\\
& (\hat{\gamma}^{ij} \> \hat{P}_{kl})|_{t} \>\Psi= -i \> \hbar \> \gamma^{ij} \>\frac{\delta \Psi}{\delta_{\Sigma_t} \gamma^{kl}},
\end{aligned}
\end{equation}

\noindent where we have made use of the expression in the first line,
\begin{equation} \label{4GR-FunctionalDerivativeDeltas}
\begin{aligned}
\frac{\delta \, {\gamma}^{ij}(y)}{\delta_{\Sigma_{t}} \, {\gamma}^{kl}(y^\prime)}=\delta^i_k \> \delta^j_l \> \delta^3(y-y^\prime),
\end{aligned}
\end{equation}

\noindent with \(\delta(y-y^\prime)\) being the Dirac delta function. We now write down the formal commutation relations:
\begin{equation} \label{4QG-Commutator}
\begin{aligned}
{} \left[\hat{\gamma}^{ij}(y), \hat{P}_{kl}(y^\prime) \right]=i \, \hbar \> \delta^i_k \> \delta^j_l \> \delta^3(y-y^\prime) \> \hat{\mathbb{I}}.
\end{aligned}
\end{equation}

\subsection{Hamiltonian operator and the Wheeler-DeWitt equation}
We begin by generalizing the computation (\ref{4QM-Wavefunctionqp}) to quantum general relativity:
\begin{equation} \label{4QG-DerivGravitationalPathIntegralMomMetric}
\begin{aligned}
(\hat{\gamma}^{ij} \> \hat{P}_{kl})|_{t} \>\Psi= -i \> \hbar \> \gamma^{ij} \>\frac{\delta \Psi}{\delta_{\Sigma_t} \gamma^{kl}} =\gamma^{ij}|_{\Sigma_{t}} \int \mathscr{D} [g^{\alpha \beta}] \, P_{kl} |_{\Sigma_{t}}  \, e^{\frac{i}{\hbar} S_{GR}[g^{\alpha \beta}]}=\int \mathscr{D} [g^{\alpha \beta}] \, ( \gamma^{ij} \, P_{kl}) |_{\Sigma_{t}}  \, e^{\frac{i}{\hbar} S_{GR}[g^{\alpha \beta}]}.
\end{aligned}
\end{equation}

\noindent The above expression demonstrates that quantities of the form \(( \gamma^{ij} \, P_{kl}) |_{\Sigma_{t}}=( P_{kl} \, \gamma^{ij}) |_{\Sigma_{t}}\) that appear in the integrand of path integrals will yield the operator ordering \((\hat{\gamma}^{ij} \> \hat{P}_{kl})|_{t}\). 

For path integrals containing higher factors of the momenta \(P_{ij}\), it is instructive to recall our earlier discussion of superspace \(\mathscr{S}(\Sigma)\) and its manifold extension \(\mathscr{S}_{ex}(\Sigma)\). Recall that in terms of the coordinates \(\xi^{\mathfrak{a}}\) on extended superspace \(\mathscr{S}_{ex}(\Sigma)\) (which we assume to be of countable dimension) the gravitational Hamiltonian \(H_{GR}\) is the Hamiltonian for a particle on an infinite-dimensional pseudo-Riemannian manifold. In the superspace representation, we demand that the wave function \(\Psi(\xi)\) is invariant under coordinate transformations on \(\mathscr{S}_{ex}(\Sigma)\), which suggests the following correspondence [cf. Eq. (\ref{4QM-OperatorLaplaceBeltrami})]:
\begin{equation} \label{4QG-LBCorrespondenceRule}
\int \mathscr{D}\xi \>({\mathscr{G}}^{\mathfrak{a} \mathfrak{b}} \, P_{\mathfrak{a}} \, P_{\mathfrak{b}})|_{t_2} \>  \> e^{(i/\hbar) \> S[q]}=(- i \, \hbar)^2 (\hat{\mathscr{G}}^{\mathfrak{a} \mathfrak{b}} \nabla_{\mathfrak{a}} \, \nabla_{\mathfrak{b}})|_{t_2} \, \Psi .
\end{equation}

\noindent This is the Laplace-Beltrami operator on the extended superspace \(\mathscr{S}_{ex}(\Sigma)\).\footnote{It should be mentioned that our approach differs from that suggested by DeWitt in Ref. \cite{DeWitt1967}, who proposed using a Laplace-Beltrami operator on the six-dimensional manifold coordinatized by the 3-metric \textit{components} \(\gamma_{ij}\) (as opposed to the inverse 3-metric \textit{functions} \(\gamma^{ij}(y)\)). Our approach instead uses the Laplace-Beltrami operator on the infinite-dimensional extended superspace manifold \(\mathscr{S}_{ex}(\Sigma)\).} In the superspace representation, the Hamiltonian operator \(H_{GR}\) takes the form [cf. Equation (\ref{3GR-HamiltonianSuperspaceRep})]:
\begin{equation} \label{3GR-HamiltonianOperatorSuperspaceRep}
\begin{aligned}
\hat{H}_{GR}&=\frac{\varepsilon \, \hbar^2 }{2 \kappa} \left( \sum^\infty_{\mathfrak{a}=1}\sum^\infty_{\mathfrak{b}=1} \hat{\mathscr{G}}^{\mathfrak{a} \mathfrak{b}} \, \nabla_{\mathfrak{a}} \,  \nabla_{\mathfrak{b}}\right) + \hat{\Phi},
\end{aligned}
\end{equation}

\noindent where \(\hat{\Phi}\) is the operator counterpart to the potential \(\Phi(\xi)\) in (\ref{3GR-HamiltonianPotential}).

To obtain the standard representation, we construct a map\footnote{We will later provide an example of how one might construct such a map.} \(\xi^{\mathfrak{a}}\llbracket \gamma^{ij} \rrbracket:  iRiem(\Sigma) \rightarrow \mathscr{S}_{ex}(\Sigma)\) and an inverse map\footnote{A coordinate condition on \(\gamma^{ij}(y)\) is needed in order to map an element of \(\mathscr{S}_{ex}(\Sigma)\) uniquely to an element of \(iRiem(\Sigma)\).} \(\gamma^{ij}(y,\xi):\mathscr{S}_{ex}(\Sigma) \rightarrow iRiem(\Sigma)\). 
We then make use of the chain rule
\begin{equation} \label{3GR-ChainRule}
\begin{aligned}
\frac{\delta}{\delta \gamma^{ij}} &= \sum^\infty_{\mathfrak{c}=1} \frac{\delta \xi^{\mathfrak{c}}}{\delta \gamma^{ij}} \> \frac{\partial}{\partial \xi^{\mathfrak{c}}}\\
\frac{\partial}{\partial \xi^{\mathfrak{c}}}&=\int_{\Sigma} d^3y \> \frac{\partial \gamma^{ij}(y,\xi)}{\partial \xi^{\mathfrak{c}}}\frac{\delta}{\delta \gamma^{ij}}
\end{aligned}
\end{equation}

\noindent to obtain the expression for the action of \(\hat{H}_{GR}\) on the wave functional \(\Psi=\Psi \llbracket \gamma^{ij} \rrbracket\),
\begin{equation} \label{3GR-HamiltonianOperatorMetricRep}
\begin{aligned}
\hat{H}_{GR} \, \Psi &=  \frac{1}{2 \kappa} \int_{\Sigma_t} \, \left[\varepsilon \, \hbar^2 \,  {G}^{ijkl} \, \frac{\delta}{\delta \gamma^{ij}}\frac{\delta \Psi}{\delta \gamma^{kl}} + \varepsilon \, \hbar^2 \, \mathscr{C}^{mn}(y,\xi\llbracket \gamma^{ij} \rrbracket) \, \frac{\delta \Psi}{\delta \gamma^{mn}} - {^3}\hat{R} \, \hat{\sqrt{|\gamma|}} \, \Psi \right] d^3y,
\end{aligned}
\end{equation}

\noindent where any ``unhatted'' factors are multiplicative operators, and we define
\begin{equation} \label{3GR-CFactor}
\begin{aligned}
\mathscr{C}^{mn}(y,\xi\llbracket \gamma^{ij} \rrbracket):=\sum^\infty_{\mathfrak{a}=1}\sum^\infty_{\mathfrak{b}=1} \sum^\infty_{\mathfrak{c}=1} \mathscr{G}^{\mathfrak{a} \mathfrak{b}} \left[ \int_{\Sigma_t} \left( \frac{\partial \gamma^{pq}(z,\xi)}{\partial \xi^{\mathfrak{a}}} \frac{\delta \xi^{\mathfrak{c}}}{\delta \gamma^{pq}} \frac{\partial^2 \, \gamma^{mn}(y,\xi)}{\partial \xi^{\mathfrak{c}} \, \partial \xi^{\mathfrak{b}}}\right) d^3z - \Gamma^{\mathfrak{c}}_{\mathfrak{a}{\mathfrak{b}}}\,\frac{\partial \gamma^{mn}}{\partial \xi^{\mathfrak{c}}} \right],
\end{aligned}
\end{equation}

\noindent where \(\Gamma^{\mathfrak{c}}_{\mathfrak{a}{\mathfrak{b}}}\) is given by (\ref{3GR-SuperspaceConnection}). The meaning of the second functional derivative (\ref{3GR-HamiltonianOperatorMetricRep}) will be discussed in an upcoming paper.

We now return to Eq. (\ref{4QG-VariationGravitationalPathIntegralHam}), which becomes [using (\ref{4QG-HamiltonianDensityGaugeFixed})]
\begin{equation} \label{4QG-HamiltonianOperatorCond2}
\begin{aligned}
\hat{H}_{GR} \, \Psi&=\frac{1}{2\kappa} \int \mathscr{D} [g^{\alpha \beta}] \, \biggl( \int_{\Sigma_t} \biggl[K_{ij} \> K^{ij} - K^2  - {^3}{R}\biggr]\sqrt{\gamma} \, d^3 y \biggr) e^{\frac{i}{\hbar} S_{GR}[g^{\alpha \beta}]}.
\end{aligned}
\end{equation}

\noindent The Hamiltonian constraint forms part of the Einstein field equations, and with the choice \(\alpha=1\) and \(\beta^i=0\) (valid on the boundary \(\partial \textbf{U}\)), we may write
\begin{equation} \label{4QG-HamconstraintEFE}
\begin{aligned}
2 \, {R}_{00} -  R \> g_{00}=K_{ij} \> K^{ij} - K^2 - {^3}{R}.
\end{aligned}
\end{equation}

\noindent If we recall the earlier result (\ref{4QG-VariationGravitationalPathIntegralBulk}), namely, that the vacuum Einstein field equations are satisfied within the path integral,
\begin{equation} \label{4QG-VariationGravitationalPathIntegralBulk2}
\begin{aligned}
\int \mathscr{D} [g^{\alpha \beta}] \, \left({R}_{\mu \nu} -  \frac{1}{2} \>R \> g_{\mu \nu} \right) \sqrt{|g|}\, e^{\frac{i}{\hbar} S_{GR}[g^{\alpha \beta}]}=0,
\end{aligned}
\end{equation}

\noindent we find that (\ref{4QG-HamconstraintEFE}) and (\ref{4QG-VariationGravitationalPathIntegralBulk2}) imply that the right-hand side of (\ref{4QG-HamiltonianOperatorCond2}) must vanish, so
\begin{equation} \label{4QG-WheelerDeWittEquationInt}
\begin{aligned}
\hat{H}_{GR} \, \Psi=0 \>\>\>\>\> \Rightarrow \>\>\>\>\> \frac{1}{2 \kappa}  \int_{\Sigma_t} \, \left[\varepsilon \, \hbar^2 \, {G}^{ijkl} \, \frac{\delta}{\delta \gamma^{ij}}\frac{\delta \Psi}{\delta \gamma^{kl}} + \varepsilon \, \hbar^2 \, \mathscr{C}^{mn}(y,\xi\llbracket \gamma^{ij} \rrbracket) \, \frac{\delta \Psi}{\delta \gamma^{mn}} - {^3}\hat{R} \, \hat{\sqrt{|\gamma|}} \, \Psi \right] d^3y=0.
\end{aligned}
\end{equation}

\noindent Since the Hamiltonian operator determines the time evolution [recall (\ref{4QG-VariationGravitationalPathIntegralHam})], this suggests that the wave functional \(\Psi\) must be independent of the time parameter \(t\):
\begin{equation} \label{4QG-WavefunctionalStatic}
\begin{aligned}
\frac{\partial\Psi}{\partial t} =0.
\end{aligned}
\end{equation}

\noindent The time independence of the wave functional presents both conceptual and technical difficulties. This is known as the ``problem of time'' in the canonical formulation of quantum general relativity. A full discussion of the conceptual and technical aspects of the problem of time in the canonical theory (in particular for spacetimes without spatial boundary) is beyond the scope of this paper---we refer the reader to the review articles \cite{Anderson2012,*Kuchar1992,*Isham1992} and also Ref. \cite{Kiefer2012QG} for a brief overview. Our results explicitly demonstrate that the problem of time also persists in the path integral approach to quantum general relativity in the case of spacetimes without a spatial boundary. Other derivations of the Wheeler-DeWitt equation, for instance the derivation in Ref. \cite{HartleHawking1983}, might also lead one to infer that the problem of time persists in the path integral formulation, but we believe that our derivation is more explicit. In path integral approaches to quantum gravity, the problem of time may actually be more severe, at least in the case of spacetimes without a spatial boundary; it has been pointed out \cite{Kiefer2001,Kiefer2012QG} that the gravitational path integral effectively contains an integral over the time parameter via the functional integral over the lapse function of the metric (or in our case, the functional integral over \(g^{00}\)). We will later demonstrate that one can gain some degree of control over this issue in the case of spacetimes with a spatial boundary.

We stress that (\ref{4QG-WheelerDeWittEquationInt}) is not what is usually referred to as the Wheeler-DeWitt equation; the Wheeler-DeWitt equation \cite{Wheeler1968}, \cite{DeWitt1967} is the \textit{local} counterpart to (\ref{4QG-WheelerDeWittEquationInt}), which one may infer from the \(0\)\(0\) component of (\ref{4QG-VariationGravitationalPathIntegralBulk2}),
\begin{equation} \label{4QG-WheelerDeWittEquationPrec}
\begin{aligned}
2 \kappa \, \hat{\mathscr{H}}_{gf} \, \Psi = \varepsilon \, \hbar^2 \, {G}^{ijkl} \, \frac{\delta}{\delta \gamma^{ij}}\frac{\delta \Psi}{\delta \gamma^{kl}} + \varepsilon \, \hbar^2 \, \mathscr{C}^{mn}(y,\xi\llbracket \gamma^{ij} \rrbracket) \, \frac{\delta \Psi}{\delta \gamma^{mn}} - {^3}\hat{R} \, \hat{\sqrt{|\gamma|}} \, \Psi = 0,
\end{aligned}
\end{equation}

\noindent where \(\mathscr{C}^{mn}(y,\xi\llbracket \gamma^{ij} \rrbracket) \) is defined in Eq. (\ref{3GR-CFactor}). Again, the meaning of the second functional derivative in (\ref{4QG-WheelerDeWittEquationPrec}) will be discussed in an upcoming paper. Note that this equation is local in \(y^i\), rather than a global one over the whole of \(\Sigma_t\). This form of the Wheeler-DeWitt equation differs from the form presented by DeWitt in Ref. \cite{DeWitt1967}; in particular, DeWitt argues that one can ignore operator ordering issues by requiring that multiple functional derivatives \(\delta/\delta \gamma^{ij}\) vanish when acting on the same spacetime point. It was later argued by Tsamis and Woodard that one can indeed ignore operator ordering issues when employing dimensional regularization---see Ref. \cite{TsamisWoodard1987}. In such cases, we may drop the term containing \(\mathscr{C}^{mn}(y,\xi\llbracket \gamma^{ij} \rrbracket)\).

Before proceeding, we note that the \(0\)\(i\) component of (\ref{4QG-VariationGravitationalPathIntegralBulk2}) leads to the momentum constraint
\begin{equation} \label{4QG-MomentumConstraint}
\begin{aligned}
\gamma^{ik} \, D_k \left(\frac{2 \kappa}{\sqrt{|\gamma|}} \, \frac{\delta \Psi}{\delta \gamma^{ij}}\right)=0,
\end{aligned}
\end{equation}

\noindent which follows from (\ref{3GR-MomConstraint}). One may infer from the general Weiss variation (\ref{3GR-GravitationalActionVariationFullWeiss}) that this constraint corresponds to the invariance of the wave functional \(\Psi \llbracket \gamma^{ij}\rrbracket\) under coordinate transformations on \(\Sigma_t\); in other words, the functional \(\Psi \llbracket \gamma^{ij}\rrbracket\) must be covariant. Earlier, we required the measure \(\mathscr{D}[g^{\alpha \beta}]\) to be defined so that the wave function has this property, but if one were to seek functionals \(\Psi \llbracket \gamma^{ij} \rrbracket\) that solve the Wheeler-DeWitt Eq. (\ref{4QG-WheelerDeWittEquationPrec}) in the absence of a path integral definition, one should check that the solutions \(\Psi \llbracket \gamma^{ij} \rrbracket\) also satisfy the constraint (\ref{4QG-MomentumConstraint}).

\section{Quantum general relativity for spacetimes with spatial boundary}
\subsection{Functional integral and its variation}\label{WdWSpatial}
We now consider what happens when we consider path integrals for regions of spacetime \(\textbf{W}\) with a spatial boundary, as described by Fig. \ref{3GR-BoundaryNotation}. In particular, we begin with a functional integral of the form [with \(S_{GR,\textbf{B}}\) defined in (\ref{3GR-GravitationalActionSpatialBoundaries})],
\begin{equation} \label{4QG-GravitationalPathIntegralSpB}
\begin{aligned}
\mathcal{K} \llbracket h^{ij}_{I}; h^{ij}_{F}; q^{ab} \rrbracket=\int \mathscr{D} [g^{\alpha \beta}] \, e^{\frac{i}{\hbar} S_{GR,\textbf{B}}[g^{\alpha \beta}]},
\end{aligned}
\end{equation}

\noindent where \(h^{ij}_{I}\) is the induced metric on \(\Sigma_I\), \(h^{ij}_{F}\) is the induced metric on \(\Sigma_F\), and \(q^{ab}\) is the induced metric on \(\textbf{B}\). Here, we view the path integral \(\mathcal{K} \llbracket h^{ij}_{I}; h^{ij}_{F}; q^{ab} \rrbracket\) as a transition amplitude for the gravitational field, subject to the spatial boundary condition that the induced metric on \(\textbf{B}\) is given by \(q^{ab}\). This viewpoint is essentially the same as that of the general boundary formulation of quantum field theory \cite{Oeckl2003,ConradyRovelli2004}, in which the state of a quantum field is specified on the (connected) boundary of a compact region of spacetime---as pointed out in Refs. \cite{Oeckl2003b,Oeckl2008}, this approach avoids the problem of time, since time evolution is specified by the boundary conditions. In the remainder of this section, we will demonstrate how the boundary conditions specify time evolution.

For any functional of the form \(\mathcal{K} \llbracket h^{ij}_{I}; h^{ij}_{F}; q^{ab} \rrbracket\), one may write
\begin{equation} \label{4QG-GravitationalPathIntegralSpBVar2a}
\begin{aligned}
\delta \mathcal{K}=\int_{\Sigma_I}\,\frac{\delta \mathcal{K}}{\delta h^{ij}_{I}}\> \Delta h^{ij}_I \> d^3 y+\int_{\textbf{B}}\, \frac{\delta \mathcal{K}}{\delta  q^{ab}} \> \Delta q^{ab}  \> d^3 y+\int_{\Sigma_F}\,\frac{\delta \mathcal{K}}{\delta  h^{ij}_{F}}\> \Delta h^{ij}_{F} \> d^3 y ,
\end{aligned}
\end{equation}

\noindent where \(\Delta h^{ij}_{I}\), \(\Delta q^{ab}\), and \(\Delta h^{ij}_{F}\) are variations of the respective functions \(h^{ij}_{I}\), \(q^{ab}\), and \(h^{ij}_{F}\). The methods outlined in the preceding section may be used to obtain path integral expressions for the functional derivatives. To do so, we consider boundary conditions \(\{h^{ij}_{I}; h^{ij}_{F}; q^{ab}\}\) that admit solutions of the vacuum Einstein field equations. The variation of the path integral \(\mathcal{K} \llbracket h^{ij}_{I}; h^{ij}_{F}; q^{ab} \rrbracket\) takes the form
\begin{equation} \label{4QG-GravitationalPathIntegralSpBVar}
\begin{aligned}
\delta \mathcal{K} =\frac{i}{\hbar}\int \mathscr{D} [g^{\alpha \beta}] \, (\delta S_{GR,\textbf{B}}) \, e^{\frac{i}{\hbar} S_{GR,\textbf{B}}[g^{\alpha \beta}]}.
\end{aligned}
\end{equation}

\noindent As before, we require the variations to satisfy \(\delta x^\mu|_{{\textbf{S}}_I}=0\), \(\delta x^\mu|_{{\textbf{S}}_F}=0\), \(\delta g_{\mu \nu}|_{{\textbf{S}}_I}=0\), and \(\delta g_{\mu \nu}|_{{\textbf{S}}_F}=0\), and again, we also require \(\langle n_I,n_{\textbf{B}} \rangle|_{{\textbf{S}}_I}\) and \(\langle n_F,n_{\textbf{B}} \rangle|_{{\textbf{S}}_F}\) to be held fixed under the variations so that the variation of the Hayward term vanishes. From the arguments in the preceding section, we may infer that the Einstein field equations (rescaled by a factor of the volume element \(\sqrt{|g|}\)) are satisfied inside the path integral so that
\begin{equation} \label{4QG-GravitationalPathIntegralSpBEFE}
\begin{aligned}
\int \mathscr{D} [g^{\alpha \beta}] \,\left({R}_{\mu \nu} -  \frac{1}{2} \>R \> g_{\mu \nu} \right) \sqrt{|g|}  \, e^{\frac{i}{\hbar} S_{GR,\textbf{B}}[g^{\alpha \beta}]} = 0.
\end{aligned}
\end{equation}

\noindent It follows that the Hamiltonian and momentum constraints (which are formed from the Einstein field equations) are satisfied inside the path integral. Particularly important are the Hamiltonian constraints, which take the following form (again, recall that an underline denotes quantities defined on \(\textbf{B}\))
\begin{equation} \label{4QG-GravitationalPathIntegralHam}
\begin{aligned}
&\int \mathscr{D} [g^{\alpha \beta}] \, \mathscr{H}_{I}  \, e^{\frac{i}{\hbar} S_{GR,\textbf{B}}[g^{\alpha \beta}]}=-\frac{1}{2\kappa}\int \mathscr{D} [g^{\alpha \beta}] \, \biggl[\left({^3}{R}+ (K^2-K_{ij} \> K^{ij}) \right)\sqrt{h}  \biggr]_{\Sigma_I} \, e^{\frac{i}{\hbar} S_{GR,\textbf{B}}[g^{\alpha \beta}]}=0\\
&\int \mathscr{D} [g^{\alpha \beta}] \, \mathscr{H}_{\textbf{B}} \, e^{\frac{i}{\hbar} S_{GR,\textbf{B}}[g^{\alpha \beta}]}=-\frac{1}{2\kappa} \int \mathscr{D} [g^{\alpha \beta}] \, \biggl[\left({^3}\underline{R}-(\underline{K}^2-\underline{K}_{ab} \> \underline{K}^{ab}) \right)\sqrt{q}\biggr]_{\textbf{B}} \, e^{\frac{i}{\hbar} S_{GR,\textbf{B}}[g^{\alpha \beta}]}=0\\
&\int \mathscr{D} [g^{\alpha \beta}] \, \mathscr{H}_{F}  \, e^{\frac{i}{\hbar} S_{GR,\textbf{B}}[g^{\alpha \beta}]}=-\frac{1}{2\kappa}\int \mathscr{D} [g^{\alpha \beta}] \, \biggl[\left({^3}{R}+ (K^2-K_{ij} \> K^{ij})\right)\sqrt{h} \biggr]_{\Sigma_F}  \, e^{\frac{i}{\hbar} S_{GR,\textbf{B}}[g^{\alpha \beta}]}=0.
\end{aligned}
\end{equation}

\noindent where the Hamiltonian densities \(\mathscr{H}_{I}\), \(\mathscr{H}_{\textbf{B}}\) and \(\mathscr{H}_{F}\) are given in Eq. (\ref{3GR-HamiltonianDensitySpatial}). Using (\ref{3GR-GravitationalActionVariationFullWeiss2b}) and the above constraints, we find that the variation of the transition amplitude \(\mathcal{K}\) takes the form (again, we stress that \(I\) and \(F\) are labels, not indices to be summed over)
\begin{equation} \label{4QG-GravitationalPathIntegralSpBVar2b}
\begin{aligned}
\delta \mathcal{K} &=\frac{i}{\hbar}\int \mathscr{D} [g^{\alpha \beta}] \, \biggl(\int_{\Sigma_I}\, P^I_{ij} \> \Delta h_I^{ij}  \> d^3 y+\int_{\textbf{B}}\, \underline{P}_{ab} \> \Delta q^{ab}  \> d^3 y + \int_{\Sigma_F}\, P^F_{ij} \> \Delta h^{ij}_{F} \> d^3 y \biggr) \, e^{\frac{i}{\hbar} S_{GR,\textbf{B}}[g^{\alpha \beta}]},
\end{aligned}
\end{equation}

\noindent where [cf. Eq. (\ref{3GR-MomentumDensitySpatial})]
\begin{equation} \label{4QG-MomentumDensitySpatial}
\begin{aligned}
{P}^I_{ij}&:=-\frac{1}{2 \kappa}\,({K}_{ij}-{K} \> h^I_{ij})  \, \sqrt{h_I}\\
\underline{P}_{ab}&:=\frac{1}{2 \kappa}\,(\underline{K}_{ab}-\underline{K} \> q_{ab})  \, \sqrt{|q|}\\
{P}^F_{ij}&:=-\frac{1}{2 \kappa}\,({K}_{ij}-{K} \> h^F_{ij})  \, \sqrt{h_F}.
\end{aligned}
\end{equation}

\noindent We now compare (\ref{4QG-GravitationalPathIntegralSpBVar2b}) with the general expression (\ref{4QG-GravitationalPathIntegralSpBVar2a}) for the variation \(\delta \mathcal{K}\) to establish a relationship between the following operators and their corresponding path integral expressions:
\begin{equation} \label{4QG-GravitationalPathIntegralSpBDerivatives}
\begin{aligned}
&\hat{P}^I_{ij} \, \mathcal{K} \llbracket h^{ij}_{I}; h^{ij}_{F}; q^{ab} \rrbracket=- i \, \hbar \, \frac{\delta \mathcal{K}}{\delta  h^{ij}_{I}}=\int \mathscr{D} [g^{\alpha \beta}] \,P^I_{ij}\, e^{\frac{i}{\hbar} S_{GR,\textbf{B}}[g^{\alpha \beta}]}\\
&\hat{P}^{\textbf{B}}_{ab} \, \mathcal{K} \llbracket h^{ij}_{I}; h^{ij}_{F}; q^{ab} \rrbracket= -i \, \hbar \, \frac{\delta \mathcal{K}}{\delta  q^{ab}}=\int \mathscr{D} [g^{\alpha \beta}] \,\underline{P}_{ab}\, e^{\frac{i}{\hbar} S_{GR,\textbf{B}}[g^{\alpha \beta}]}\\
&\hat{P}^F_{ij} \, \mathcal{K} \llbracket h^{ij}_{I}; h^{ij}_{F}; q^{ab} \rrbracket= -i \, \hbar \, \frac{\delta \mathcal{K}}{\delta h^{ij}_{F}}=\int \mathscr{D} [g^{\alpha \beta}] \,P^F_{ij}\, e^{\frac{i}{\hbar} S_{GR,\textbf{B}}[g^{\alpha \beta}]}.
\end{aligned}
\end{equation}

\noindent We may define the following operators:
\begin{equation} \label{4QG-GravitationalPathIntegralSpBMetricOps}
\begin{aligned}
&\hat{h}_I^{ij} \, \mathcal{K} \llbracket h^{ij}_{I}; h^{ij}_{F}; q^{ab} \rrbracket={h}^{ij}_I \, \mathcal{K} \llbracket h^{ij}_{I}; h^{ij}_{F}; q^{ab} \rrbracket\\
&\hat{q}^{ij} \, \mathcal{K} \llbracket h^{ij}_{I}; h^{ij}_{F}; q^{ab} \rrbracket={q}^{ij} \, \mathcal{K} \llbracket h^{ij}_{I}; h^{ij}_{F}; q^{ab} \rrbracket\\
&\hat{h}^{ij}_F \, \mathcal{K} \llbracket h^{ij}_{I}; h^{ij}_{F}; q^{ab} \rrbracket={h}^{ij}_F \, \mathcal{K} \llbracket h^{ij}_{I}; h^{ij}_{F}; q^{ab} \rrbracket.
\end{aligned}
\end{equation}

Equations (\ref{4QG-GravitationalPathIntegralHam}) then demand that the functional \(\mathcal{K} \llbracket h^{ij}_{I}; h^{ij}_{F}; q^{ab} \rrbracket\) must satisfy the following functional differential equations, which we collectively call the \textit{extended Wheeler-DeWitt equations}, since we extend the formalism to spatial boundaries (for simplicity, we suppress the argument of \(\mathcal{K} \llbracket h^{ij}_{I}; h^{ij}_{F}; q^{ab} \rrbracket\)),
\begin{equation} \label{4QG-WdWSpB}
\begin{aligned}
2 \kappa \, \hat{\mathscr{H}}_{I} \,\mathcal{K} & =- \hbar^2 \, G^{ijkl}_I\, \frac{\delta}{\delta h_I^{ij}}\frac{\delta \mathcal{K}}{\delta h_I^{kl}} - \hbar^2 \, \mathscr{C}^{mn}(y,\xi\llbracket h_I^{ij} \rrbracket) \, \frac{\delta \mathcal{K}}{\delta h_I^{mn}} - {^3}\hat{R}_I \, \hat{\sqrt{h_I}} \, \mathcal{K}=0\\
2 \kappa \, \hat{\mathscr{H}}_{\textbf{B}} \, \mathcal{K} & = \hbar^2 \,\underline{G}^{abcd} \, \frac{\delta}{\delta q^{ab}}\frac{\delta \mathcal{K}}{\delta q^{cd}}+\hbar^2 \, \underline{\mathscr{C}}^{ab}(y,\underline{\xi}\llbracket q^{cd} \rrbracket) \, \frac{\delta \mathcal{K}}{\delta q^{ab}} - {^3}\hat{R}_{\textbf{B}} \, \hat{\sqrt{|q|}} \, \mathcal{K}=0\\
2 \kappa \, \hat{\mathscr{H}}_{F} \, \mathcal{K} & = -\hbar^2 \, G^{ijkl}_F \, \frac{\delta}{\delta h_F^{ij}}\frac{\delta \mathcal{K}}{\delta h_F^{kl}} - \hbar^2 \, \mathscr{C}^{mn}(y,\xi\llbracket h_F^{ij} \rrbracket) \, \frac{\delta \mathcal{K}}{\delta h_F^{mn}} - {^3}\hat{R}_F \, \hat{\sqrt{h_F}} \, \mathcal{K}=0,
\end{aligned}
\end{equation}

\noindent where \(\underline{\mathscr{C}}^{ab}(y,\underline{\xi}\llbracket q^{cd} \rrbracket)\) is given by the formula (\ref{3GR-CFactor}) with all quantities replaced by underlined quantities and \(\gamma^{ij}\) replaced with \(q^{ab}\). Equations (\ref{4QG-WdWSpB}) form the extended Wheeler-DeWitt equations for spacetimes with a spatial boundary. To ensure that \(\mathcal{K} =\mathcal{K} \llbracket h^{ij}_{I}; h^{ij}_{F}; q^{ab} \rrbracket\) is invariant under coordinate transformations on the surfaces \(\Sigma_I\), \(\Sigma_F\), and \(\textbf{B}\), the extended Wheeler-DeWitt equations must be supplemented by the momentum constraints:
\begin{equation} \label{4QG-MomentumConstraintSpB}
\begin{aligned}
&h_I^{ik} \, D_k \left(\frac{2 \kappa}{\sqrt{h_I}} \, \frac{\delta \mathcal{K}}{\delta h_I^{ij}}\right)=0\\
&q^{ac} \, \underline{D}_c \left(\frac{2 \kappa}{\sqrt{|q|}} \, \frac{\delta \mathcal{K}}{\delta q^{ab}}\right)=0\\
&h_F^{ik} \, D_k \left(\frac{2 \kappa}{\sqrt{h_F}} \, \frac{\delta \mathcal{K}}{\delta h_F^{ij}}\right)=0.
\end{aligned}
\end{equation}

\noindent The problem of finding the dependence of the transition amplitude \(\mathcal{K}\) on \(h^{ij}_{I}\), \(h^{ij}_{F}\), and \(q^{ab}\) amounts to solving these functional differential equations. 

It should be cautioned that Eqs. (\ref{4QG-WdWSpB}) and (\ref{4QG-MomentumConstraintSpB}) may not be sufficient to fully determine the transition amplitude \(\mathcal{K}\). So far, we have ignored the dependence of the transition amplitude on the 2-surfaces \(\textbf{S}_I\) and \(\textbf{S}_F\), in particular, the role of the rapidity angles \(\eta_I\) and \(\eta_F\) [defined in (\ref{3GR-HaywardIntegrand})]. A simple example suggests that meaningful transition amplitudes must also depend on the rapidity angles \(\eta_I\) and \(\eta_F\). Consider the Schwarzschild solution in Painlev\'e-Gullstrand coordinates \cite{Poisson,ChoquetBruhatIntroGR,Gourgolhoun3+1}:
\begin{equation}\label{1-PGLineElement}
\begin{aligned}
ds^2&=-\left(1-\frac{2 G M}{r}\right)dT^2+2 \> \sqrt{\frac{2 G M}{r}} \> dT \> dr+dr^2 + r^2 (d\theta^2+\text{sin}^2\theta \> d\phi^2).
\end{aligned}
\end{equation}

\noindent One may construct a boundary in the manner of Fig. \ref{3GR-BoundaryNotation} from cylindrical hypersurfaces of constant \(r\) and (flat) hypersurfaces of constant \(T\) that is isometrically embeddable in Minkowski spacetime; the induced metrics \(h^{ij}_{I}\), \(h^{ij}_{F}\), and \(q^{ab}\) will be identical for Schwarzschild spacetime and Minkowski spacetime. Meaningful transition amplitudes therefore cannot be solely dependent on \(h^{ij}_{I}\), \(h^{ij}_{F}\), and \(q^{ab}\). On the other hand, note that due to the cross-terms in (\ref{1-PGLineElement}), the rapidity angles \(\eta_I\) and \(\eta_F\) (\ref{3GR-HaywardIntegrand}) at the corners of the cylinder (the 2-surfaces \(\textbf{S}_I\) and \(\textbf{S}_F\)) differ between Schwarzschild spacetime and Minkowski spacetime. This suggests that meaningful transition amplitudes must depend on the rapidity angles \(\eta_I\) and \(\eta_F\) at the 2-surfaces \(\textbf{S}_I\) and \(\textbf{S}_F\) (in addition to the induced metric on \(\textbf{S}_I\) and \(\textbf{S}_F\)).\footnote{Upon closer analysis, one may note that, while the surfaces of constant \(T\) are flat, they each contain the point \(r=0\), which corresponds to the location of the Schwarzschild singularity. The questions of whether a cylindrical boundary constructed from a surface of constant \(T\) is admissible, and whether one must include an additional boundary surface formed from excising curvature singularities from the manifold (which may require prior assumptions about the structure of singularities and topology of the manifold), are left for future analysis.} To obtain the dependence of the transition amplitude \(\mathcal{K}\) on the rapidity angles, one must also consider variations of the corner terms \(S_{C}\) in the gravitational action and the variations due to displacements of the 2-surfaces \(\textbf{S}_I\) and \(\textbf{S}_F\). This will be left for future work.

\subsection{Spatial boundaries and time evolution}
One obstacle to addressing the question of time in the case of spacetimes without spatial boundaries is the fact that the path integral involves summing over all functional forms (up to diffeomorphisms) for the spacetime metric \(g^{\mu \nu}\). Consider a timelike geodesic in \(\textbf{U}\), defined by initial conditions (the initial position and 4-velocity) at a point on \(\Sigma_I\). The proper time along the geodesic segment between the hypersurfaces \(\Sigma_I\) and \(\Sigma_F\) depends on the spacetime metric \(g^{\mu \nu}\). The path integral over \(g^{\mu \nu}\) therefore prevents one from unambiguously establishing a notion for the time elapsed between \(\Sigma_I\) and \(\Sigma_F\). 

In spacetimes with spatial boundary, the aforementioned obstacle may be used to provide a resolution for the problem of time. Recall that the spatial boundary conditions for the metric tensor \(g^{\mu \nu}\) are provided by the induced metric \(q^{ab}\), which in turn specifies the geometry of the boundary surface \(\textbf{B}\). Now, consider a timelike geodesic on \(\textbf{B}\) (a geodesic with respect to the boundary metric \(q_{ab}\)), defined by an initial 3-velocity and an initial starting point on \(\textbf{S}_I\). The induced metric \(q^{ab}\) determines the elapsed proper time along the geodesic segment between the surfaces \(\textbf{S}_I\) to \(\textbf{S}_F\). The boundary metric \(q^{ab}\) therefore provides a measure of time elapsed between the hypersurfaces \(\Sigma_I\) and \(\Sigma_F\). This suggests the following view: time should not be treated as a local parameter but as a property of the geometry of the spatial boundary \(\textbf{B}\).\footnote{One might recognize the relationship between this viewpoint and Mach's principle (see Ref. \cite{KhouryParikh2009} for a modern discussion of Mach's principle in GR), noting that the geometry of the spatial boundary \(\textbf{B}\) is determined by the spacetime geometry outside of the region \(\textbf{W}\), which in turn depends on the matter configuration outside of \(\textbf{W}\). The definition of time as a property of the boundary geometry for \(\textbf{B}\) suggests that time is fundamentally nonlocal quantity that depends on the spacetime geometry and matter configurations in distant regions.} 

More explicitly, one may imagine describing time evolution as a displacement of the surface \(\Sigma_F\) in the future time direction, with a corresponding ``stretch'' of the boundary \(\textbf{B}\) \cite{BrownYork1993}.\footnote{The results and arguments presented in this section are equivalent to those of Brown and York in Ref. \cite{BrownYork1993}.} One may note, however, that the displacement of the surface \(\Sigma_F\) by way of \(\delta x^\mu\) is a diffeomorphism and is mathematically indistinguishable from an infinitesimal coordinate transformation. On the other hand, a stretch of the boundary \(\textbf{B}\) in the timelike direction that corresponds to an increase in the proper time of timelike geodesics cannot be represented as a change in coordinates.\footnote{To better see that displacements of the boundary are not sufficient to describe \textit{physical} time evolution, consider the action \(S_{GR,\textbf{B}}[g^{\mu \nu}]\) evaluated on a vacuum solution (in which the Ricci scalar takes the value \(R=0\)) with a cylindrical boundary as described in Fig. \ref{3GR-BoundaryNotation}. Displace a portion of the boundary \(\Sigma_F\) in the normal direction by a function \(\delta x^\mu(y)\) that vanishes on \(\textbf{S}_F\). At the same time, perform an infinitesimal variation of the bulk metric tensor \(g^{\mu \nu}\) under the condition that the boundary metric \(\gamma^{ij}\) is held fixed when the boundary is displaced; in other words, we perform a variation such that \(\delta x^\mu \neq 0\) and \(\Delta \gamma^{ij}=0\). Equation (\ref{3GR-GravitationalActionVariationFullWeiss2b}) suggests that the variation \(\delta S_{GR,\textbf{B}}[g^{\mu \nu}]\) vanishes by virtue of the Hamiltonian constraint \(\mathscr{H}_I=0\) [where \(\mathscr{H}_I\) is defined in (\ref{3GR-HamiltonianDensitySpatial})] on the boundary. Displacements of the boundary surface (in particular the \(t_F\) surface \(\Sigma_F\)) have no effect on the value of the action \(S_{GR,\textbf{B}}[g^{\mu \nu}]\) (but stretching of the boundary surface \(\textbf{B}\) does).} In fact, we may ignore boundary displacements altogether and characterize the stretch in the boundary by changing the components of the boundary metric \(q_{ab}\).

To illustrate how the components of the boundary metric \(q_{ab}\) may be used to stretch the boundary in the timelike direction, we consider a boundary \(\textbf{B}\) with coordinates \((t,\theta,\phi)\), with the domain
\begin{equation} \label{3GR-BoundaryDomain}
\begin{aligned}
& t_1 < t < t_2\\
& 0 < \theta < \pi\\
& 0 < \phi < 2 \pi.
\end{aligned}
\end{equation}

\noindent We place a metric \(q_{ab}\) on \(\textbf{B}\), which admits the following line element:
\begin{equation} \label{3GR-BoundaryLineElementExample}
\begin{aligned}
ds^2=q_{ab} \, dy^a \, dy^b=- \underline{\alpha}^2 \, dt^2+ r^2 \left( d\theta^2+\sin^2 \theta \, d\phi^2 \right).
\end{aligned}
\end{equation}

\noindent The boundary metric \(q_{ab}\) corresponds to a particular set of boundary conditions on \(\textbf{B}\). If \(\underline{\alpha}\) and \(r\) are constants, then the proper time along a timelike geodesic on \(\textbf{B}\) defined by \(\theta=\text{constant}\) and \(\phi=\text{constant}\) is given by the expression \(T=\underline{\alpha} \, (t_2-t_1)\). The physical stretching of the boundary \(\textbf{B}\) in the timelike direction corresponds to an increase in the value of \(\underline{\alpha}\). Now, one might note that changes in the value of \(\underline{\alpha}\) may also be interpreted as a rescaling of the time coordinate \(t\), which in turn may be interpreted as a coordinate transformation. However, what distinguishes our construction from a coordinate transformation is that the domain (\ref{3GR-BoundaryDomain}) of the coordinates on the manifold \(\textbf{B}\) is held fixed---in particular, the coordinate values \(t_1< t <t_2\) that define the boundary \(\textbf{B}\) are held fixed. The only thing we change is the component of the metric \(q_{00}=-\underline{\alpha}^2\).

To see how such a stretch in the boundary affects the action (and by extension the path integral \(\mathcal{K} \llbracket h^{ij}_{I}; h^{ij}_{F}; q^{ab} \rrbracket\)), consider the 3-volume for \(\textbf{B}\), which may be written as
\begin{equation} \label{3GR-BoundaryExampleVolume}
\begin{aligned}
V(\textbf{B})=4 \, \pi \int_{t_1}^{t_2} \underline{\alpha} \,  r^2 \, dt= 4 \, \pi \, r^2 \, T ,
\end{aligned}
\end{equation}

\noindent where, again, \(T=\underline{\alpha} \, (t_2-t_1)\) is the proper time of a geodesic defined by defined by \(\theta=\text{constant}\) and \(\phi=\text{constant}\). For the boundary metric in (\ref{3GR-BoundaryLineElementExample}), the formula above establishes a relationship between the proper time of certain observers on the boundary and the 3-volume \(V(\textbf{B})\). Note that the presence of the Gibbons-Hawking-York boundary term in the gravitational action \(S_{GR,\textbf{B}}\) ensures that, even for vacuum solutions of the Einstein field equations, the gravitational action \(S_{GR,\textbf{B}}\) has a nonvanishing value. The Gibbons-Hawking-York boundary term for a spacetime with the spatial boundary \(\textbf{B}\) with line element (\ref{3GR-BoundaryLineElementExample}) will ultimately depend on \(\underline{\alpha}\), since the 3-volume (in particular, the volume element for \(\textbf{B}\)) depends on \(\underline{\alpha}\). The path integral \(\mathcal{K} \llbracket h^{ij}_{I}; h^{ij}_{F}; q^{ab} \rrbracket\) for the same spatial boundary \(\textbf{B}\) will in turn depend on \(\underline{\alpha}\) as well; time evolution for spacetimes with spatial boundaries admitting a line element of the form (\ref{3GR-BoundaryLineElementExample}) corresponds to an increase in \(\underline{\alpha}\).

For more general boundary geometries, we may consider a boundary metric \(q_{ab}\) written in a form adapted to the foliation induced by the coordinate \(t\). In particular, we construct the line element on \(\textbf{B}\)
\begin{equation} \label{3GR-BoundaryLineElementGeneral}
\begin{aligned}
ds^2=q_{ab} \, dy^a \, dy^b=- (\underline{\alpha}^2+ \sigma_{A B}\, \underline{\beta}^A \, \underline{\beta}^B) \, dt^2 + \sigma_{A B} \, \underline{\beta}^A \, dz^B \, dt + \sigma_{A B}\, dz^A \, dz^B ,
\end{aligned}
\end{equation}

\noindent where \(z^1=\theta\), \(z^2=\phi\), \(\sigma_{AB}\) is the induced metric on (two-dimensional) surfaces of constant \(t\), and the quantities \(\underline{\alpha}\) and \(\underline{\beta}^A\) are in general functions of \(t\) and \(z^A\). A reader familiar with the Arnowitt-Deser-Misner(ADM) formalism \cite{MTW,ADM62} will recognize the above as the ADM decomposition for the boundary line element \(ds^2=q_{ab} \, dy^a  \,dy^b\). As before, a stretch in the boundary corresponds to an increase in the value of boundary ``lapse'' function \(\underline{\alpha}\). That an increase in \(\underline{\alpha}\) corresponds to a stretch in the boundary can be seen in the expression for the 3-volume of \(\textbf{B}\)
\begin{equation} \label{3GR-BoundaryVolumeGeneral}
\begin{aligned}
V(\textbf{B})=\int_{\textbf{B}} \sqrt{|q|} \, d^3y=\int_{t_1}^{t_2} \int_{0}^{2 \pi} \int_{0}^{\pi} \underline{\alpha} \, \sqrt{|\det(\sigma_{AB})|} \, d\theta \, d\phi \, dt ,
\end{aligned}
\end{equation}

\noindent where we have made use of the expression \( \sqrt{|q|}=\underline{\alpha} \, \sqrt{|\det(\sigma_{AB})|}\). From the above expression for the 3-volume \(V(\textbf{B})\), it is clear that an overall increase in the value of \(\underline{\alpha}\) will increase the 3-volume of the boundary \(\textbf{B}\).

The notion that the lapse function \(\underline{\alpha}\) governs time evolution has been explored before in Ref. \cite{BrownYork1993}; variations of the action with respect to \(\underline{\alpha}\) were used to obtain the Brown-York quasilocal energy. We may obtain the Brown-York quasilocal energy (up to a reference term) from the expression (\ref{3GR-GravitationalActionVariationFullWeiss2b}) for \(\delta S_{GR,\textbf{B}}\) by first noting that the components \(q^{ab}\) depend on \(\underline{\alpha}\) in the following way:
\begin{equation} \label{3GR-InverseBoundaryMetricADM}
\begin{aligned}
q^{00}&=-\underline{\alpha}^{-2}\\
q^{0A}&= \underline{\alpha}^{-2} \> \underline{\beta}^A\\
q^{AB}&=-\underline{\alpha}^{-2} \> \underline{\beta}^A \> \underline{\beta}^{B}+\sigma^{AB} .
\end{aligned}
\end{equation}

\noindent The Brown-York quasilocal energy\footnote{We use the definition given in Eq. 4.3 of Ref. \cite{BrownYork1993}.} \(1/\kappa \int_{\textbf{S}}  \, \bar{n}_a \bar{n}_b (\underline{K}_{ab} - \underline{K}\, q_{ab}) \sqrt{|\sigma_{AB}|} \, d^2z\) is given by the integral (over a spacelike 2-surface \(\textbf{S}\subset \textbf{B}\)) of the functional derivative
\begin{equation} \label{3GR-GravitationalActionDerivativeq00}
\begin{aligned}
\frac{\underline{\alpha}}{\sqrt{|q|}}\frac{\delta S_{GR,\textbf{B}}}{\delta \underline{\alpha} }=2\frac{1}{\underline{\alpha}^2\sqrt{|q|}}\left(\frac{\delta S_{GR,\textbf{B}}}{\delta q^{00}}-2\frac{\delta S_{GR,\textbf{B}}}{\delta q^{0A}} \underline{\beta}^A+\frac{\delta S_{GR,\textbf{B}}}{\delta q^{AB}} \underline{\beta}^A \, \underline{\beta}^B\right)=\frac{1}{\kappa} \, \bar{n}^a \, \bar{n}^b \left(\underline{K}_{ab}-\underline{K} \> q_{ab}\right),
\end{aligned}
\end{equation}

\noindent where \([\bar{n}^a]:=(1/\underline{\alpha},-\beta^A/\underline{\alpha})\) are vectors tangent to \(\textbf{B}\) that have unit norm and are normal to surfaces of constant \(t\). Following Ref. \cite{BrownYork1993}, one may obtain similar expressions for momentumlike and stresslike quantities by performing variations with respect to \(\beta^A\) and \(\sigma^{AB}\).

Finally, the above expression for the variation may be used to obtain the change in the transition amplitude with respect to changes in \(\underline{\alpha}\),
\begin{equation} \label{3GR-TransitionAmplitudeVariationLapse}
\begin{aligned}
\hat{\underline{\alpha}}^3 \frac{\delta \mathcal{K}}{\delta \underline{\alpha} }= -\left(\hat{P}^{\textbf{B}}_{00}- \hat{\underline{\beta}}^A \, \hat{P}^{\textbf{B}}_{0 A}+\hat{\underline{\beta}}^A \, \hat{\underline{\beta}}^B \, \hat{P}^{\textbf{B}}_{ab}\right) \mathcal{K} \llbracket h^{ij}_{I}; h^{ij}_{F}; q^{ab} \rrbracket,
\end{aligned}
\end{equation}

\noindent where the operators \(\hat{\underline{\alpha}}\) and \(\hat{\underline{\beta}}^A\) pick out the value of the quantities \({\underline{\alpha}}=-|q^{00}|^{-1/2}\) and \(\hat{\underline{\beta}}^A=-q^{0A}/q^{00}\) from the inverse boundary metric \(q^{ab}\):
\begin{equation} \label{3GR-BoundaryLapseShiftOperators}
\begin{aligned}
\hat{\underline{\alpha}} \>\mathcal{K} \llbracket h^{ij}_{I}; h^{ij}_{F}; q^{ab} \rrbracket=-|q^{00}|^{-1/2}\,\mathcal{K} \llbracket h^{ij}_{I}; h^{ij}_{F}; q^{ab} \rrbracket\\
\hat{\underline{\beta}}^A\, \mathcal{K} \llbracket h^{ij}_{I}; h^{ij}_{F}; q^{ab} \rrbracket=-q^{0A}/q^{00}\,\mathcal{K} \llbracket h^{ij}_{I}; h^{ij}_{F}; q^{ab} \rrbracket .
\end{aligned}
\end{equation}

\noindent Equation (\ref{3GR-TransitionAmplitudeVariationLapse}) is a form\footnote{Our result (\ref{3GR-BoundaryLapseShiftOperators}) uses a slightly more general form of the Brown-York quasilocal energy than that of the boundary Schr{\"o}dinger equation in Ref. \cite{HaywardWong1992}.} of the boundary ``Schr{\"o}dinger equation'' in Ref. \cite{HaywardWong1992}. Though it is tempting to regard (\ref{3GR-TransitionAmplitudeVariationLapse}) as \textit{the} Schr{\"o}dinger equation for the transition amplitude \(\mathcal{K} \llbracket h^{ij}_{I}; h^{ij}_{F}; q^{ab} \rrbracket\), it is in fact a kinematical expression as it merely expresses one functional derivative of \(\mathcal{K} \llbracket h^{ij}_{I}; h^{ij}_{F}; q^{ab} \rrbracket\) in terms of other functional derivatives [more pointedly, (\ref{3GR-TransitionAmplitudeVariationLapse}) is simply a statement of the chain rule for functional derivatives]. Equation (\ref{3GR-TransitionAmplitudeVariationLapse}) therefore does not determine the dynamics for the theory.  The dependence of the transition amplitude on spatial boundary conditions comes from the requirement that the transition amplitude \(\mathcal{K} \llbracket h^{ij}_{I}; h^{ij}_{F}; q^{ab} \rrbracket\) satisfies the extended Wheeler-DeWitt equations (\ref{4QG-WdWSpB}) and the momentum constraints (\ref{4QG-MomentumConstraintSpB}).
The Wheeler-DeWitt equations (\ref{4QG-WdWSpB}) and the momentum constraints (\ref{4QG-MomentumConstraintSpB}) determine the functional dependence of the transition amplitudes \(\mathcal{K} \llbracket h^{ij}_{I}; h^{ij}_{F}; q^{ab} \rrbracket\) on \(h^{ij}_{I}\), \(h^{ij}_{F}\), and \(q^{ab}\). Once the solutions to (\ref{4QG-WdWSpB}) and (\ref{4QG-MomentumConstraintSpB}) are found, Eq. (\ref{3GR-TransitionAmplitudeVariationLapse}) may then be used to extract the explicit time dependence for the transition amplitude \(\mathcal{K} \llbracket h^{ij}_{I}; h^{ij}_{F}; q^{ab} \rrbracket\).

\subsection{Dependence of transition amplitude on spatial boundary conditions}
We briefly describe how the formalism presented in this paper might be used to compute the dependence of transition amplitudes on spatial boundary conditions. If the spatial boundary \(\textbf{B}\) has the cylindrical topology \(\bar{\mathbb{R}} \times \mathbb{S}^2\) (recall that \(\bar{\mathbb{R}}\) is a compact subset of the real line \(\mathbb{R}\)) and the appropriate boundary conditions are imposed on \(\textbf{S}_I\) and \(\textbf{S}_F\), one may decompose the induced metric in the manner
\begin{equation} \label{F-Decomposition}
q^{ab}(t,\theta,\phi)=q_0^{ab}(t,\theta,\phi)+\sum_{l m n} \left(A^m_{l \,n} \sin(\pi \, n \, t/T) + B^m_{l \,n} \sin(\pi \, n \, t/T) \right) Y_l^m(\theta \phi) ,
\end{equation}

\noindent where \(A^m_{l \,n}\) and \(B^m_{l \,n}\) are constant coefficients, \(T\) is the length of the real line segment \(\bar{\mathbb{R}}\), and \(q_0^{ab}(t,\theta,\phi)\) are the components of inverse metric for the cylindrical line element:
\begin{equation} \label{F-BoundaryLineElementCylindrical}
ds^2=- dt^2+ r^2 \left( d\theta^2+\sin^2 \theta \, d\phi^2 \right).
\end{equation}

\noindent Since (\ref{F-Decomposition}) is formed from a complete, orthogonal basis for functions on \(\bar{\mathbb{R}} \times \mathbb{S}^2\), the coefficients \(A^m_{l \,n}\) and \(B^m_{l \,n}\) may be used to coordinatize\footnote{One should be aware that two sets of values for the coefficients may correspond to the same point in \(\underline{\mathscr{S}}_{ex}(\textbf{B})\), as they may describe equivalent 3-geometries that differ by a coordinate transformation.} the manifold \(\underline{\mathscr{S}}_{ex}(\textbf{B})\). This may be used to convert the spatial boundary Wheeler-DeWitt Eq. (\ref{4QG-WdWSpB}) from a functional differential equation to an infinite number of PDEs on an infinite-dimensional manifold (note that the spatial boundary Wheeler-DeWitt equation is a {local} equation in the metric basis---it is a function of $y\in \textbf{B}$). One may truncate the series (\ref{F-Decomposition}) to obtain a finite number of PDEs on a finite-dimensional manifold; if these can be solved,\footnote{Since the transition amplitude satisfies more than one equation, the system is overdetermined, and there is a danger that truncation may lead to an inconsistent set of equations.} one can obtain the dependence of the transition amplitudes on spatial boundary conditions up to truncation errors. The remaining Wheeler-DeWitt equations (\ref{4QG-WdWSpB}) may be solved in a similar manner to obtain the dependence of the transition amplitude on \(h^{ij}_I\) and \(h^{ij}_F\).

A potentially tricky aspect of this procedure is to obtain the functional \(\xi^{\mathfrak{c}}\llbracket q^{ab}\rrbracket\), which is a map from the space of 3-metrics \(q^{ab}(y)\) to the coordinates \(\xi^{\mathfrak{c}}\) on \(\underline{\mathscr{S}}_{ex}(\textbf{B})\). The orthogonality of the basis functions, which we write as \(e_{\mathfrak{c}}(y)\) may be exploited to obtain the functionals \(\xi^{\mathfrak{c}}\llbracket  q^{ab} \rrbracket\) from the functions \(q^{ab}(y)\). Using the shorthand \(\bar{q}(y)\) to represent \(q^{ab}(y)\) (we suppress indices for simplicity), we may write \(\bar{q}(y)=\xi^{\mathfrak{c}} \, e_{\mathfrak{c}}(y)\). The functional \(\xi^{\mathfrak{c}}\llbracket  \bar{q} \rrbracket\) may then be written as (no sum over \(\mathfrak{c}\))
\begin{equation} \label{3GR-Functional}
\begin{aligned}
\xi^{\mathfrak{c}}\llbracket  \bar{q} \rrbracket=\frac{\int_{\textbf{B}} \bar{q}(y) \,  e_{\mathfrak{c}}(y) \, d^3y}{\int_{\textbf{B}} \, e_{\mathfrak{c}}(y) \, e_{\mathfrak{c}}(y) \, d^3y}.
\end{aligned}
\end{equation}

\noindent This may be varied in order to obtain an expression for \({\delta \xi^{\mathfrak{c}}}/{\delta q^{ab}}\) in the chain rule formula (\ref{3GR-ChainRule}). In turn, one may use the chain rule (\ref{3GR-ChainRule}) to rewrite the Wheeler-DeWitt equation on \(\textbf{B}\) (\ref{4QG-WdWSpB}) in terms of coordinates on superspace \(\underline{\mathscr{S}}_{ex}(\textbf{B})\).\footnote{It may be necessary to multiply the Wheeler-DeWitt equation by a factor of the inverse metric \(q^{ij}(y)\); the dependence on the coordinate \(y\) acts as part of a ``free index'', and the factor of the inverse metric \(q^{ij}(y)\) allows one to convert the Wheeler-DeWitt equation to a set of PDEs indexed by the superspace coordinate index \(\mathfrak{c}\).}

\section{Summary and Future Work}

We claim that when spatial boundaries are included transition amplitudes in quantum general relativity satisfy the extended Wheeler-DeWitt Eq. (\ref{4QG-WdWSpB}) and the momentum constraint (\ref{4QG-MomentumConstraintSpB}), which in turn determine the dependence of the transition amplitude on the components of the boundary metric \(q^{ab}\), which constitute boundary conditions for the metric tensor \(g^{\mu \nu}\) at the spatial boundary \(\textbf{B}\). We have argued that time evolution for transition amplitudes corresponds to a stretching of the spatial boundary \(\textbf{B}\) in the timelike direction, which in turn may be described by changes to the components of the boundary metric \(q^{ab}\). In short, we find that \textit{spatial} boundary conditions determine time evolution in quantum general relativity; our results formalize, validate, and sharpen the general idea\cite{Oeckl2003b,Oeckl2008} that time evolution is determined by boundary conditions for a connected boundary of a compact spacetime.

As argued at the end of Sec. \ref{WdWSpatial}, the formalism presented in this paper is not sufficient, even at a formal level, to fully determine the transition amplitude \(\mathcal{K}\); further development of the formalism is needed. Critically important is the dependence of \(\mathcal{K}\) on the rapidity angles \(\eta_I\) and \(\eta_F\) and the induced metric at the 2-surfaces \(\textbf{S}_I\) and \(\textbf{S}_F\). To do this, one must obtain the variation of \(S_{GR,\textbf{B}}\) when variations in the rapidity angles \(\eta_I\) and \(\eta_F\), variations in the induced metric on \(\textbf{S}_I\) and \(\textbf{S}_F\), and the displacement of the surfaces \(\textbf{S}_I\) and \(\textbf{S}_F\) are included.

We have briefly described a way to convert the functional Wheeler-DeWitt equation to a set of PDEs on an infinite-dimensional manifold; one may truncate the function space in order to obtain PDEs on a \textit{finite}-dimensional manifold. The next step is to write down the explicit expressions for the resulting PDEs and to study their general properties. This may be attempted for both the case of spatially compact 3-geometries {without} a boundary and compact regions of spacetime {with} a spatial boundary.

\begin{acknowledgments}
This article is based on the dissertation work of J. C. Feng. We thank Mr. Mark Selover, Prof. E. C. G. Sudarshan and Prof. G. Bhamathi for their comments and encouragement. J. C. Feng also thanks Prof. Austin Gleeson, Prof. Philip J. Morrison, Prof. Richard D. Hazeltine, and Prof. Robert E. Gompf for their guidance and service as members of his dissertation committee. This work was partially supported by the National Science Foundation under Grant No. PHY-1620610.
\end{acknowledgments}

\bibliography{bibStd,bibGRQG,bibGRADM,bibSptop,bibOurs}

\appendix
\section{The Weiss variation of the gravitational action}\label{Appendix-WGR}
As stated earlier, the full justification for Eq. (\ref{3GR-GravitationalActionVariationFullWeiss}) is given in Ref. \cite{FengMatzner2017a}. However, for the benefit of the reader, we present a partial justification for (\ref{3GR-GravitationalActionVariationFullWeiss}) in this Appendix, valid for \textit{spacelike} boundary surfaces (\(\varepsilon=-1\)). In particular, we discuss here how one might \textit{infer} Eq. (\ref{3GR-GravitationalActionVariationFullWeiss}) from results in the literature (excluding Ref. \cite{FengMatzner2017a}) and the Weiss variation formula (\ref{3M-WeissVariation}) for mechanical systems, which we rewrite here [setting \(\delta q^i(t)=\lambda \, \eta^i(t)\)]:
\begin{equation} \label{A-WeissVariation}
\begin{aligned}
\delta S &=\int^{t_2}_{t_1} \left(\frac{\partial L}{\partial q^i} - \frac{d}{dt} \left(\frac{\partial L}{\partial \dot{q}^i} \right)\right) \delta q^i(t) \> dt + \left( p_i \> \Delta q^i - H \, \Delta t\right) \biggr |^{t_2}_{t_1}.
\end{aligned}
\end{equation}

We begin by recalling that for the gravitational action in (\ref{3GR-GravitationalAction}) the quantity to be held fixed on the boundary \(\partial\textbf{U}\) in Hamilton's principle is the induced metric \(\gamma_{ij}\) \cite{York1986}. Equivalently, we may instead require the inverse metric $\gamma^{ij}$ to be held fixed on the boundary in Hamilton's principle. This suggests that a natural choice for the configuration variables in gravity is the inverse induced metric \(\gamma^{ij}(x)\). From Ref. \cite{York1986}, the variation of the gravitational action (\ref{3GR-GravitationalAction}) is (excluding boundary displacements)
\begin{equation} \label{A-GravitationalActionVariationYork}
\begin{aligned}
\delta S_{GR}&=\frac{1}{2 \kappa}\int_{\textbf{U}}  \, {G}_{\mu \nu} \,  \delta g^{\mu \nu} \sqrt{|g|} \,  d^4 x+ \frac{ \varepsilon}{2\kappa}\int_{\partial \textbf{U}} \, (K_{ij} - K \> \gamma_{ij}) \, \delta \gamma^{ij} \, \sqrt{|\gamma|} \, d^3 y.
\end{aligned}
\end{equation}

\noindent The reader should be aware that the definitions for \(K_{ij}\) and \(K\) in Ref. \cite{York1986} differ from the ones here by a sign. Also, the boundary variation in Ref. \cite{York1986} is expressed in terms of \(\delta \gamma_{ij}\); we use the first-order expression \(\delta \gamma_{ij}=-\gamma_{ia} \, \gamma_{jb} \, \delta \gamma^{ij}\) to obtain (\ref{A-GravitationalActionVariationYork}). Equation (\ref{A-GravitationalActionVariationYork}) suggests the definition for the conjugate field momentum tensor
\begin{equation} \label{A-ConjugateFieldMomentumTensor}
p_{ij}:=K_{ij} - K \> \gamma_{ij},
\end{equation}

\noindent which is equivalent to the definition in (\ref{3GR-ConjugateMomentumTensor}). 

It is well known (see Refs. \cite{Poisson,Gourgolhoun3+1}) that for spacetimes {without} a spatial boundary the Hamiltonian density for the gravitational field may be written in the form \(\mathscr{H}=\alpha \, \mathscr{H}_{gf} + \beta_i \, \mathscr{C}^i\), where \(\alpha\) and \(\beta^i\) are the respective lapse function and shift vector of the ADM formalism \cite{ADM62}, \(\mathscr{H}_{gf}\) is gauge-fixed Hamiltonian density of (\ref{3GR-HamiltonianDensityGaugeFixed}), and \(\mathscr{C}^i\) is
\begin{equation} \label{A-MomentumConstraintVector}
\mathscr{C}^i:=-\frac{1}{\kappa}D_{j} p^{ij}\, \sqrt{|\gamma|}
\end{equation}

\noindent Explicitly, the gravitational Hamiltonian on a boundaryless spacelike hypersurface \(\Sigma_t\) is \cite{Poisson,Gourgolhoun3+1}
\begin{equation} \label{A-HamiltonianGR}
H_t=\frac{1}{2\kappa}\int_{\Sigma_t}\left[ \alpha \, (K_{ij} \> K^{ij} - K^2 - {^3}{R} ) - 2\, \beta_i\,D_j (K^{ij} - K \> \gamma^{ij})\right] \sqrt{|\gamma|} \, d^3y.
\end{equation}

\noindent The basis vector \(\partial /\partial t\) has components \(t^\mu=\delta^\mu_0\) and may be decomposed in the manner
\begin{equation} \label{A-TimeDecomp}
t^\mu=\varepsilon \, \alpha\, n^\mu + \beta^\mu ,
\end{equation}

\noindent where \(\alpha\) satisfies \(\alpha=t^\mu \, n_\mu\) and \(\beta^\mu=\gamma^\mu {_\nu} \, t^\nu\) are the components of the shift vector in the bulk coordinate basis.\footnote{A word of caution: while \(\beta^0=0\), the lowered component \(\beta_0\neq 0\). In fact, it is not too difficult to show that \(\beta_0\) is the norm of $\beta^i$.} Since \(\beta_i=\gamma_{ij} \, \beta^j\), its bulk-basis counterpart is \(\gamma_{\mu \nu}\beta^\nu=\gamma_{\mu \nu} \, t^\nu\). We may therefore rewrite the Hamiltonian (\ref{A-HamiltonianGR}) in the bulk basis
\begin{equation} \label{A-HamiltonianGRBB1}
H_t=\frac{1}{2\kappa}\int_{\Sigma_t}\left[n_\mu \, t^\mu \, (K_{ij} \> K^{ij} - K^2 - {^3}{R} ) - 2\, \gamma_{\mu \nu} \, t^\mu \,D_\rho (K^{\nu \rho} - K \> \gamma^{\nu \rho})\right] \sqrt{|\gamma|} \, d^3y,
\end{equation}

\noindent which we rewrite as
\begin{equation} \label{A-HamiltonianGRBB2}
H_t=\frac{1}{2\kappa}\int_{\Sigma_t}\left[n_\mu \, (K_{ij} \> K^{ij} - K^2 - {^3}{R} ) - 2\, \gamma_{\mu \beta} \,D_\alpha p^{\alpha \beta} \right] t^\mu \, \sqrt{|\gamma|} \, d^3y.
\end{equation}

Given formulas (\ref{A-WeissVariation}), (\ref{A-GravitationalActionVariationYork}) and (\ref{A-HamiltonianGRBB2}), one may infer that for \(\partial \textbf{U}=\Sigma_{t_1} \cup \Sigma_{t_2}\) (with \(\Sigma_{t_1}\) and \(\Sigma_{t_2}\) spacelike and boundaryless) the Weiss variation takes the form
\begin{equation} \label{A-GravitationalActionVariationWeissInf}
\begin{aligned}
\delta S_{GR}&=\frac{1}{2 \kappa}\int_{\textbf{U}}  \, {G}_{\mu \nu} \,  \delta g^{\mu \nu} \sqrt{|g|} \,  d^4 x - \frac{1}{2\kappa}\int_{\partial \textbf{U}} \, p_{ij} \, \Delta \gamma^{ij} \, \sqrt{|\gamma|} \, d^3 y+\left. \left(H_{t} \, \Delta t \right) \right|_{t_1}^{t_2}\\
\end{aligned}
\end{equation}

\noindent where \(\Delta t\) (assumed to be infinitesimal and constant over the boundary) is the amount by which the boundary surface is displaced in the coordinate \(t\). We identify the displacement vector \(\delta x^\mu= t^\mu \, \Delta t\) and write
\begin{equation} \label{A-GravitationalActionVariationWeiss2}
\begin{aligned}
\delta S_{GR}
&=\frac{1}{2 \kappa}\int_{\textbf{U}}  \, {G}_{\mu \nu} \,  \delta g^{\mu \nu} \sqrt{|g|} \,  d^4 x - \frac{1}{2\kappa}\int_{\partial \textbf{U}} \,\left( p_{ij} \, \Delta \gamma^{ij} +\left[n_\mu \, (K_{ij} \> K^{ij} - K^2 - {^3}{R} ) - 2\, \gamma_{\mu \beta} \,D_\alpha p^{\alpha \beta} \right] \delta x^\mu \right) \sqrt{|\gamma|} \, d^3 y.
\end{aligned}
\end{equation}

\noindent Since (\ref{A-GravitationalActionVariationWeiss2}) applies for any choice of time coordinate \(t\) with spacelike hypersurfaces, it can be used to describe general boundary displacements, provided that the boundary surfaces are spacelike. Equation (\ref{A-GravitationalActionVariationWeiss2}) is therefore equivalent to (\ref{3GR-GravitationalActionVariationFullWeiss}) for spacelike boundary surfaces (\(\varepsilon=-1\)). For timelike boundary surfaces, we refer the reader to \cite{FengMatzner2017a}.

\end{document}